\pdfoutput=1
\documentclass[11pt]{elsarticle}

\usepackage{graphicx}
\usepackage{amssymb,xfrac}
\usepackage{amsthm,amsmath,dsfont}
\usepackage{thmtools}
\allowdisplaybreaks

\usepackage{lineno}

\usepackage[colorlinks=true, citecolor=blue]{hyperref}
\usepackage[algoruled,titlenotnumbered]{algorithm2e}
\usepackage{algorithmic}
\usepackage{wrapfig}
\usepackage{natbib}
\usepackage{comment}
\usepackage{multirow}
\usepackage{amsmath,bm}
\usepackage{graphicx}
\usepackage[lofdepth,lotdepth]{subfig}
\usepackage[framemethod=tikz]{mdframed}

\usepackage{float}
\usepackage{subfig}
\usepackage{wrapfig}
\usepackage{framed,xcolor}
\usepackage{newfloat}
\usepackage{mathpazo}
\usepackage{amsthm}
\usepackage{thmtools}
\usepackage{stmaryrd}
\usepackage{mathtools}
\usepackage[symbol]{footmisc}
\usepackage{times}

\usepackage[colorinlistoftodos]{todonotes}

\usepackage{soul}

\usepackage{enumitem}

\newcommand{\RR}{\mathbb{R}}

\newcommand{\EE}{\mathbb{E}}

\newcommand{\PP}{\mathbb{P}}
\newcommand{\GX}{\mathcal{G}}
\newcommand{\AX}{\mathcal{A}}
\newcommand{\LX}{\mathcal{L}}

\newcommand{\NX}{\mathcal{N}}
\newcommand{\dd}{\mathrm{d}}

\newcommand{\MM}{\mathcal{M}}
\newcommand{\SX}{\mathcal{S}}
\newcommand{\mE}{\mathcal{E}}

\newcommand{\red}[1]{\textcolor{red}{#1}}

\newcommand{\green}[1]{\textcolor[rgb]{0,1,0}{#1}}

\newcommand{\orange}[1]{\textcolor[rgb]{1,0.5,0}{#1}}

\newcommand{\deeporange}[1]{\textcolor[rgb]{0.95,0.4,0.13}{#1}}

\newcommand{\deepblue}[1]{\textcolor[rgb]{0.27,0.33,0.65}{#1}}
\newcommand{\deepgreen}[1]{\textcolor[rgb]{0.07,0.61,0.28}{#1}}

\newtheoremstyle{theoremdd}
{\topsep}
{\topsep}
{\itshape}
{0pt}
{\fontfamily{cmss}\selectfont\bfseries}
{.}
{ }
{\thmname{#1}\thmnumber{ #2}\thmnote{ (#3)}}

\theoremstyle{theoremdd}
\newtheorem{theorem}{Theorem}
\newtheorem{lemma}{Lemma}
\newtheorem{proposition}{Proposition}

\newtheorem{definition}{Definition}

\renewenvironment{proof}{{\bfseries  \noindent Proof.~}}{\qed\vspace{0.5em}}

\theoremstyle{definition}
\newtheorem{example}{Example}

\usepackage{array}
\makeatletter
\newcommand{\thickhline}{%
\noalign {\ifnum 0=`}\fi \hrule height 1.5pt
\futurelet \reserved@a \@xhline
}
\newcolumntype{"}{@{\hskip\tabcolsep\vrule width 1pt\hskip\tabcolsep}}
\makeatother

\mdfdefinestyle{myStyle}{roundcorner=5pt,backgroundcolor=brown!20,linecolor=red!40!black,linewidth=1.5pt}

\journal{
}
\date{}

\biboptions{authoryear}

\usepackage{titlesec}
\titleformat*{\section}{\fontfamily{cmss}\selectfont\large\bfseries}
\titleformat*{\subsection}{\fontfamily{cmss}\selectfont\normalsize\bfseries}
\titleformat*{\subsubsection}{\fontfamily{cmss}\selectfont\normalsize}

\usepackage[left=1.0 in, right=1.0 in, top=1.0 in, bottom=1.0 in]{geometry}
\graphicspath{{./}}

\begin{document}

\begin{frontmatter}
	
	
	\title{\fontfamily{cmss}\selectfont Enlarging Stability Region of Urban Networks with Imminent Supply Prediction}
	

	
	\author[1]{Dianchao Lin}
	\author[2]{Li Li\corref{cor1}}
	\cortext[cor1]{Corresponding author, e-mail: \url{lili@fzu.edu.cn}}
	\address[1]{School of Economics and Management, Fuzhou University, Fuzhou, Fujian, China}
	\address[2]{School of Civil Engineering, Fuzhou University, Fuzhou, Fujian, China}

	{ \fontfamily{cmss}\selectfont\large\bfseries
		\begin{abstract}
			{ \normalfont\normalsize
			Stability region is a key index to characterize a dynamic processing system’s ability to handle incoming demands. It is a multidimensional space when the system has multiple OD pairs where their service rates interact. Urban traffic network is such a system. Traffic congestion appears when its demand approaches or exceeds the upper frontier of its stability region. In this decade, with the rapid development of traffic sense technology, real-time traffic operations, e.g., BackPressure (BP) control, have gained lots of research attention. Urban network's mobility could be further improved with these timely demand-responding strategies. However, most studies on real-time controls continue with traditional supply assumptions and ignore an important fact -- imminent saturation flow rate (I-SFR), i.e., the system's real-time service rate under green, is neither fixed nor given, but hard to be precisely known. It is unknown how the knowledge level of I-SFR would influence the stability region. This paper proves that knowing more accurate I-SFR can enlarge the upper frontier of the network's stability region. Furthermore, BP policy with predicted I-SFR can stabilize the network within the enlarged stability region and relieve the congestion level of the traffic network. Therefore, improving the I-SFR's prediction accuracy is meaningful for traffic operations.
			}
		\end{abstract}
	}
	
	\begin{keyword}
		Stability region \sep traffic network control \sep imminent saturation flow rate \sep BackPressure
	\end{keyword}
	
\end{frontmatter}


\setcounter{footnote}{0} 

\section{Introduction}
\label{S:intro}

The increasing urbanization process worldwide in the last decades is accompanied by significant increase in urban traffic congestion \citep{schrank2019urban,serok2022identification}. Traffic congestion appears and may spread rapidly when local demand approaches or exceeds the infrastructure's supply for a certain period of time \citep{li2019position}. To mitigate congestion and improve urban mobility, traffic control evolves from fixed-time to real-time control in the responsiveness aspect, and from isolated intersection to network control in the scale aspect \citep{webster1958traffic,  dunne1964algorithm, sheu2002stochastic, osorio2013simulation, bi2017optimal,gan2017analysis, baldi2019simulation, noaeen2021real, lin2021pay, khwais2023optimal}. Recent years, the decentralized network controls, such as Backpressure (BP) based policies \citep{wongpiromsarn2012distributed, levin2020max}, received much attention in traffic operations research since it avoids the dimension curse in centralized network control and hence can be implemented in real time in large scaled network. 

\begin{figure}[!ht] 
	\centering
	\subfloat[][Discharge with substantial interference]{\resizebox{0.36\textwidth}{!}{
			\includegraphics[width=1\textwidth]{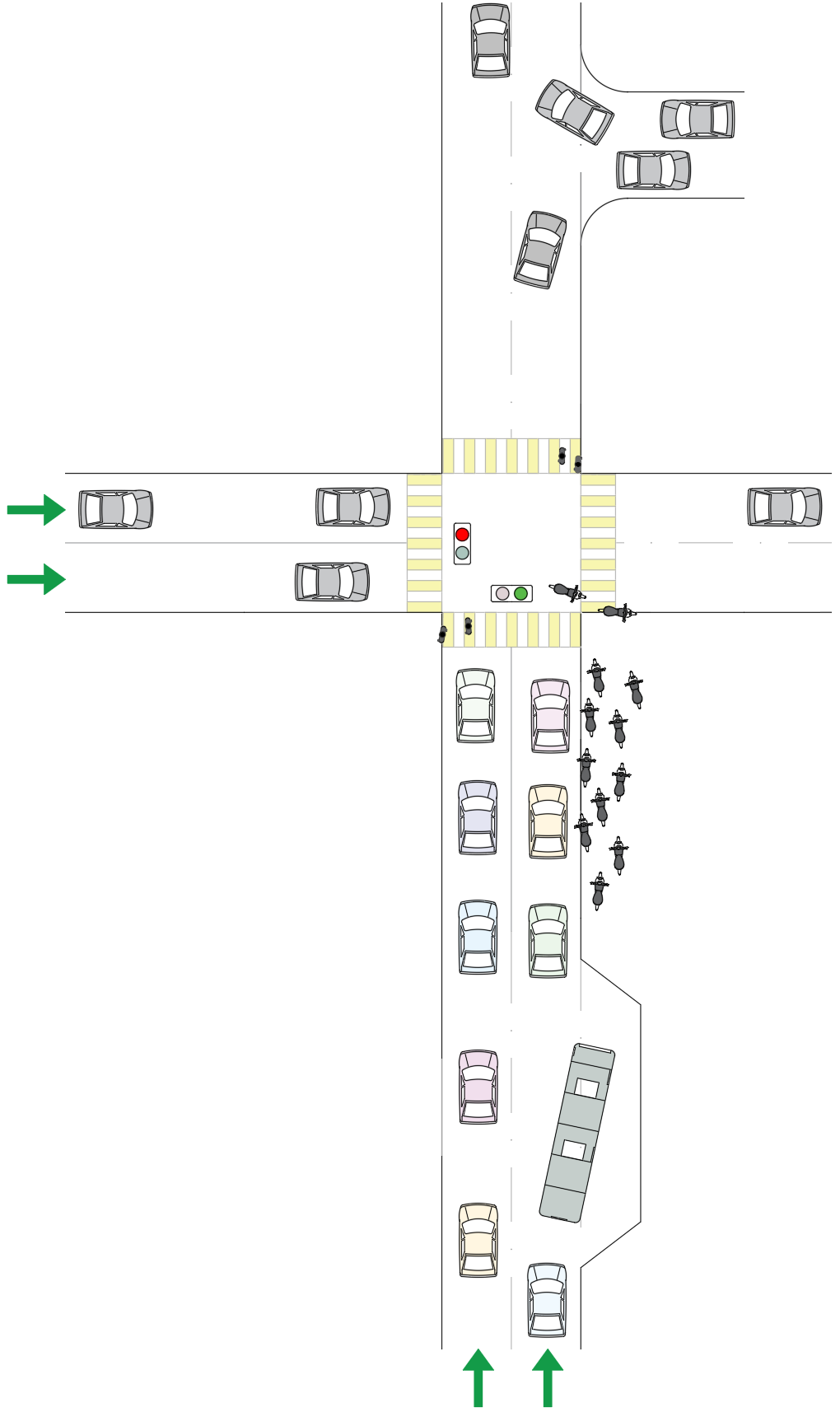}}
		\label{F:M1}} 
	~~~~~~~~~~~~~~~~~~
	\subfloat[][Discharge with little interference]{\resizebox{0.36\textwidth}{!}{
			\includegraphics[width=1\textwidth]{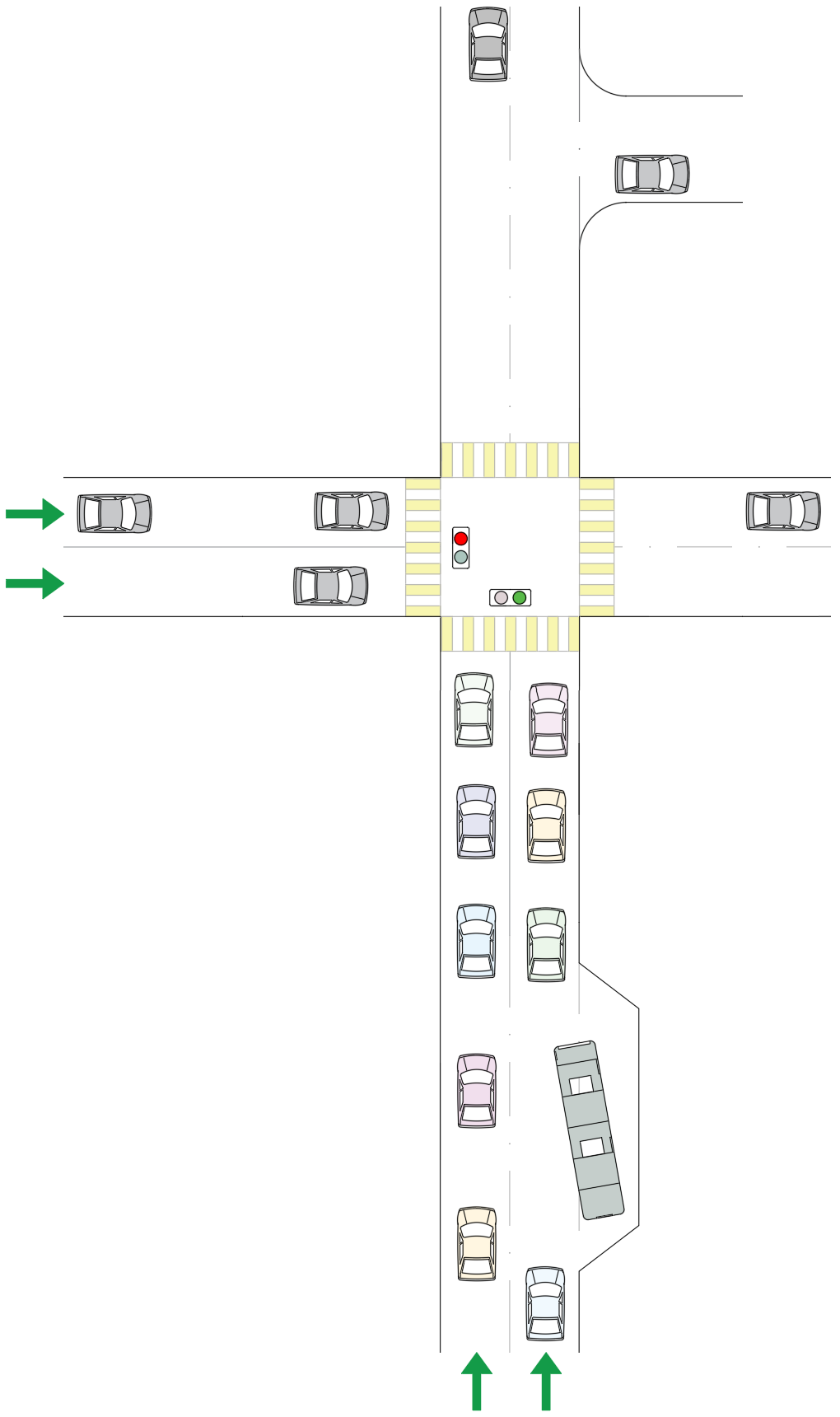}}
		\label{F:M2}}
	\caption{Two types of ``supply ability'' for the northbound automobiles in real time}
	\label{F:Sample_0}
\end{figure}

Real-time control in existing studies usually assumes a rich information environment in which demand information can be timely sensed (to some degree), and supply information is given (e.g., saturation flow rate, SFR) \citep{li2021backpressure, noaeen2021real, le2015decentralized}. However, accurate knowledge of the field traffic supply can be very hard to obtain \citep{wang2020dynamic}. \autoref{F:Sample_0} shows through an example how SFR can be influenced by complicated factors in field. Although the demands of northbound movement in \autoref{F:Sample_0}\subref{F:M1} and \autoref{F:Sample_0}\subref{F:M2} are the same, their imminent SFR (I-SFRs) next green interval should be completely different. When discharging, queues in \autoref{F:Sample_0}\subref{F:M1} will move and stop repeatedly, while vehicles in \autoref{F:Sample_0}\subref{F:M2} will pass smoothly. That is because urban road network is a complex system which serves traffic with different types (e.g., motorized vehicles, non-motorized vehicles, pedestrians) and purposes (e.g., passing, parking, and access traffic) \citep{lin2016driving}. The interactions between different traffic participants inevitably influence the automobiles' passing efficiency. Furthermore, features of the passing automobiles, such as vehicle compositions (e.g., cars, buses, heavy vehicles) and driver characteristics (e.g., aggressive or not), will also interfere with the flow efficiency \citep{hirose2022investigating, long2007driver}. 
As a result, the continuously changing traffic conditions unavoidably lead to fluctuating I-SFR, whose prediction can be challenging. If the I-SFR of each movement can be predicted more precisely, it is expected that controllers can better allocate the road priority and hence improve the network's supply ability.

The supply ability of a network is commonly described as the \emph{stability region} \citep{bramson2017proportional, dai2005maximum}, also named the capacity region or admissible demand region. It is worth noting that when researchers mention ``admissible demand'', they are actually referring to the supply, not demand. The arrival rates within the stability region can be handled by a stabilizing control under given technical background. When demand approaches or exceeds the upper frontier of the network's stability region, traffic congestion appears \citep{li2019position, rangwala2008understanding}. Therefore, increasing the upper frontier of the stability region makes the network more robust to traffic demand fluctuations and hence more reliable.
SFR of each movement can significantly influence the size of the stability region. Existing studies usually assumed SFR to be given either as a constant \citep{wei2019presslight, levin2020max} or accurate stochastic value\citep{zaidi2016back, mercader2020max}. The importance of I-SFR prediction is neglected, and it is still unknown how the uncertainty of I-SFR will affect the stability region and other network performance. 

This paper aims to investigate how the knowledge level of I-SFR affects the network's stability region, and how such influence can help relieve congestion. We take BP-based policies as the tool to verify our findings. Our main contributions are twofold: 
\begin{itemize}
	\item We prove that knowing more accurate I-SFR can enlarge the upper frontier of the network's stability region.
	\item We prove that the BP policy with predicted I-SFR can stabilize the network within the enlarged stability region. 
\end{itemize}

The proposed theory is validated through calibrated simulations and we show that congestion can be relieved. The remainder of this paper is organized as follows: \autoref{S:litr} briefly reviews the related studies about stability region, SFR uncertainty and BP control. \autoref{S:dynamics} describes the traffic dynamics model, and formally defines stability region. \autoref{S:SR} systematically deduces the relationship between stability region and SFR, and characterizes the changes in stability region with different I-SFR knowledge levels. \autoref{S:BP} proposes the BP with predicted I-SFR and demonstrates its stability. \autoref{S:simulation} shows the simulation results, and \autoref{S:Conc} concludes the paper.

\section{Related Literature}
\label{S:litr}

\subsection{Stability region}
Stability region characterizes the system's ability to carry demands. In a standard queueing system (e.g., M/M/1, M/M/c), stability region is usually a one-dimensional interval (i.e., from 0 to the total service rate), within which the arrival rate, $\lambda$, can be stably handled \citep{anton2021stability, moyal2022stability}. For a network with multiple OD pairs where their service rates interact with each other (e.g., urban network, communication network), stability region is a multi-dimensional space within which the arrival rate vector, $\bm{\lambda}$, can be accommodated \citep{dai2005maximum, bramson2017proportional}. The size of stability region can be affected by many factors, e.g., the scheduling/control policy \citep{dieker2013local, bramson2021stability} and the preciseness of used information \citep{moyal2022stability}. The stability region of a network refers to the maximum one that can be achieved by all available control policies \citep{tassiulas1992stability}, and its size can be affected by the system's given technical background. 

When demand approaches the stability region's upper frontier, the network's performance deteriorates rapidly and congestion spreads quickly \citep{rangwala2008understanding, li2019position}. Therefore, enlarging the stability region provides flexibility and resilience to the network, which allows for fluctuations and unexpected increases in demand. This is particularly important for traffic network where demand patterns can vary significantly over time (e.g., peak hours, non-peak hours) \citep{li2021backpressure, chen2019simulation}.

\subsection{Uncertainty in SFR}
In an urban traffic network, the average service rate at intersection is defined as capacity, which is determined by the SFR (given information) and effective green ratio (decision variable) \citep{hcm2010}. A movement's SFR is the number of vehicles (in saturated state) that can pass the stop line per unit time during green. Its value can be influenced by many factors such as vehicle compositions (e.g., percentage of buses, heavy vehicles) \citep{hirose2022investigating}, driver mentality (aggressive or not) \citep{long2007driver}, turning behaviors \citep{wang2010analysis}, non-motorized vehicles \citep{lin2016driving}, and pedestrians \citep{milazzo1998effect}. Those factors could vary significantly during a short period of time, and result in great fluctuations in I-SFR. 
Even when the controller can monitor the traffic conditions in real time with the help of video recognition or connected vehicle technologies, accurate prediction of I-SFR is still difficult due to its complex influencing factors \citep{wang2020dynamic, lin2023learning}. That is, the uncertainty of I-SFR information can hardly be completely removed. However, how the I-SFR's uncertainty degree will influence the network's stability region is still unknown. This paper shall fill up this research gap. 

\subsection{BP in traffic control}
BP was originally proposed in a communication network to solve the packet transmission problem \citep{tassiulas1992stability}, and was later introduced to urban traffic network by \cite{wongpiromsarn2012distributed} and \cite{varaiya2013max}. Since then, it received lots of attention and experienced continuous improvements to make it more tailed for traffic environment \citep{le2015decentralized, chai2017dynamic, li2019position, mercader2020max}. A detailed review about the development of BP in traffic operations can be found in \cite{levin2023max}. In short, BP control is 1) decentralized, 2) does not require prior knowledge of demands, and 3) can stabilize the queueing if and only if the demand is within the network's stability region, i.e., maximize the network throughput. "Pressure" of BP in traffic control is the product of demand (difference in weighted upstream and downstream queues) and supply (SFR). Controllers calculate pressure every decision interval and choose the phase with the maximum pressure to be active during this interval.

Since the SFR information in BP needs to be estimated/predicted before the vehicles are actually serviced, the used SFR may differ from the actual SFR, resulting in SFR uncertainty. However, such uncertainty is rarely considered in the literature. Most existing BP studies either assume SFR is constant and use mean SFR (M-SFR) as the constant value \citep{gregoire2014capacity, wei2019presslight, levin2020max}, or assume SFR is random but can be completely known \citep{wongpiromsarn2012distributed, zaidi2016back, mercader2020max}. \cite{varaiya2013max} realizes the randomness of SFR, but assumes the controller only knows the M-SFR. The relationship between the stability regions using different SFRs remains unknown. In case the SFR uncertainty degree changes the network's stability region, whether the BP with more/less accurate SFR still maximizes the network throughput is unknown.

\section{Problem description}
\label{S:dynamics}
\subsection{Notation}
\begin{figure}[h!]
	\centering
	\resizebox{0.5\textwidth}{!}{
		\includegraphics{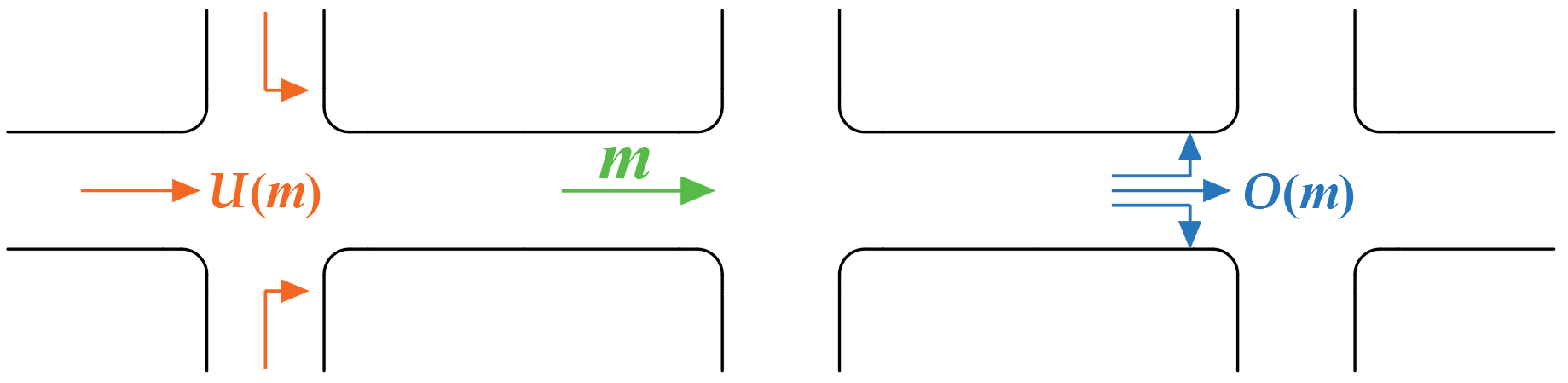}}
	\caption{Movement $m$'s upstream and downstream movements} 
	\label{F:Um_Om}
\end{figure}

Consider an urban traffic network $(\NX,\LX)$, where $\NX$ is a set of network nodes representing intersections (including an element $\emptyset$ to represent any node outside of the network), and $\LX \subset \NX \times \NX$ is a set of directed links representing road segments. Note that for links at the boundaries with start/end node outside of the network, their start/end node is represented by $\emptyset$. An allowed movement at node $n \in \NX \backslash \emptyset$ is defined as an ordered pair ($l_1$, $l_2$) such that $l_1$ belongs to the set of links terminating in $n$, and $l_2$ belongs to the links emanating from $n$. To simplify the expression of movement, we use a single letter, such as movement $m$, to represent the corresponding ordered pair. For each node $n$, let $\MM_n$ denote the set of allowed movements between its inbound and outbound links. Hence, the set of all network movements can be described as $\MM = \MM_1 \sqcup \cdots \sqcup \MM_{|\NX|-1}$, where $|\NX|$ (or $|\MM|$) is the size of $\NX$ (or $\MM$). A movement $i$ is defined as movement $m$'s upstream movement when $i$'s destination link is $m$'s origin link. Similarly, $j$ is defined as $m$'s downstream movement when $j$'s origin link is $m$'s destination link. As shown in \autoref{F:Um_Om}, we denote by $U(m)$ the set of $m$'s all upstream movements and by $O(m)$ the set of $m$'s all downstream movements. Note that if there is no upstream (or downstream) movements, $U(m)$ (or $O(m)$) is an empty set.

\begin{figure}[!ht] 
	\centering
	\subfloat[][Intersection]{\resizebox{0.485\textwidth}{!}{
			\includegraphics[width=0.5\textwidth]{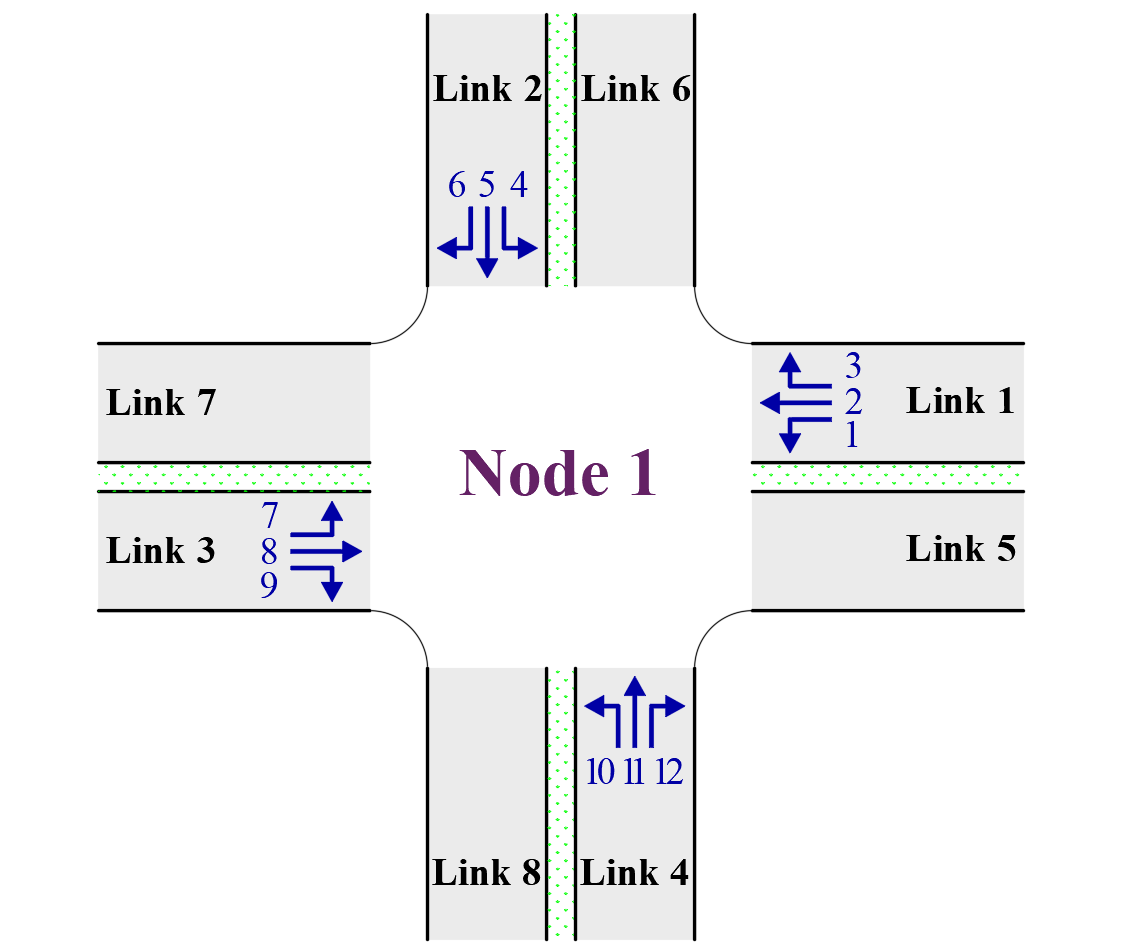}}
		\label{F:I}} 
	~~~
	\subfloat[][Phases]{\resizebox{0.36\textwidth}{!}{
			\includegraphics[width=0.4\textwidth]{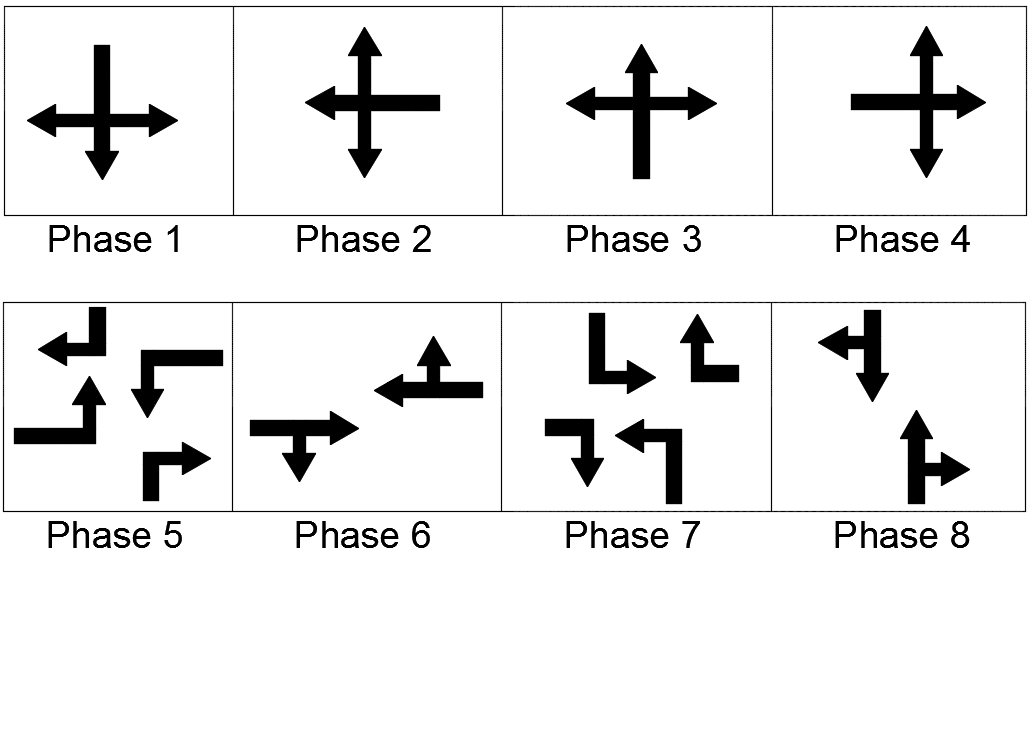}}
		\label{F:II}}
	\caption{An example of a four-legged intersection and its allowable phase set}
	\label{F:intersection_example}
\end{figure}

A signal phase consists of junction movements that do not conflict with one another. We denote by $\bm{\Phi}_n$ the set of allowable intersection phases at $n$ and by $\bm{\Phi} \equiv \bigtimes_{n \in \NX} \bm{\Phi}_n$ the set of allowable network phasing schemes. 
Example allowable phases for a four-legged intersection are shown in \autoref{F:intersection_example}. Note that if some movements at the same link share lane(s), they should be discharged simultaneously at a phase. Some phase choices in \autoref{F:intersection_example} \subref{F:II} would be unavailable in this case. The table of notations appears in \ref{Ap:notations}.

\subsection{Demand and queueing dynamics}
We discretize time into intervals of equal length and use time $t$ to refer to the beginning of interval $[t, t+1)$. For any $m \in \MM$, let $a_m(t)$ denote its exogenous arrivals during $[t, t+1)$, and let $x_m(t)$ denote its queue length at time $t$. $\boldsymbol{X}(t) = \{x_m(t)\}$ is the queue length array of all the movements within the network. Assume phase $\Phi \in \bm{\Phi}$ is selected for interval $[t, t+1)$, we let $\phi_m(t)$ denote $m$'s passing status during $[t, t+1)$, which equals 1 if passing is allowed, and 0 otherwise. Take \autoref{F:intersection_example} for example: for Node 1, when $\Phi_1 = $ "Phase 2" for $[t, t+1)$, we have that $\phi_m(t) = 1$ for $m \in \{1,2,3\}$, and $\phi_m(t) = 0$ for $m \in \{4,5,...,12\}$. Let $s_m(t)$ denote $m$'s I-SFR during $[t, t+1)$, and $r_m(t)$ denote the proportion of all vehicles from $m$'s upstream movements $U(m)$ that will directly join movement $m$ at time $t$ (with $r_m(t) = 0$ if there are no upstream vehicles). Note that for any $m \in \MM$, there is $\sum_{j \in O(m)}r_j(t) \leq 1$. 

Given a movement $m$ in the network, its demand arrivals during $[t, t+1)$ (denoted by $\lambda_m(t)$) is given by
\begin{equation}
	\lambda_m(t) = a_m(t) + r_m(t)\sum_{i \in U(m)} q_i(t,\Phi),
	\label{E:demand}
\end{equation}
where $q_i(t,\Phi)$ is real-time number of departures: 
\begin{equation}
	q_i(t,\Phi) = \min[x_i(t), s_i(t)\phi_i(t)]. 
	\label{E:q}
\end{equation}

The queueing dynamic of each movement $m$ follows
\begin{equation}
	x_m(t+1) = \max[x_m(t) - s_m(t)\phi_m(t), 0] + \lambda_m(t). 
	\label{E:Queue}
\end{equation}
Let $\lambda_m$ denote $m$'s average/expected arrival rate, and $c_m$ denote $m$'s average/expected service rate (capacity). In the traffic system, arrivals and departures satisfy the law of large numbers, and we do not distinguish their ``average value'' and ``expectation'', i.e.,
$\lambda_m =\underset{T \rightarrow \infty}{\lim} ~ \dfrac{1}{T}\sum_{t=0}^{T-1}\lambda_m(t) = \EE \{\lambda_m(t)\}$, $c_m =\underset{T \rightarrow \infty}{\lim} ~ \dfrac{1}{T}\sum_{t=0}^{T-1}s_m(t)\phi_m(t) = \EE \{s_m(t)\phi_m(t)\}$.

\begin{definition}[Rate stability]
	A movement $m$ is \textit{rate stable} if with probability one 
	\begin{equation}
		\underset{t \rightarrow \infty}{\lim} ~ \frac{x_m(t)}{t} = 0. \label{E:stable_0}
	\end{equation}
\end{definition}

\begin{proposition}
	A movement $m$ is rate stable if and only if $\lambda_m \leq c_m$. 
	\label{thm:1}
\end{proposition}
The proof of \autoref{thm:1} can be found in \cite[Theorem 2.4 a]{neely2010stochastic}. Rate stability says that a network is stable if and only if the arrival rate is no larger than the capacity for each movement. 
Considering that the actual departure rate cannot exceed the arrival rate, one can easily conclude that if a movement $m$ is stable, its average/expected departure rate would be equal to its average arrival rate in the long run, i.e., $\EE \{q_m(t)\} = \lambda_m$. 

Let $a_m$ denote the average exogenous arrival rate of $m$, and $r_m$ denote the average turning ratio of $m$. As $q_i(t)$ and $r_m(t)$ are independent, taking an expectation on both sides of \eqref{E:demand} yields
\begin{equation}
	\lambda_m = a_m + r_m \sum_{i \in U(m)}\lambda_i.
\end{equation}
As $a_m= \lambda_m  - r_m \sum_{i \in U(m)}\lambda_i$, we have 
\begin{equation} 
	\bm{a} = (\bm{I} - \bm{R})\cdot\bm{\lambda},
\end{equation}
where $\bm{a} = \left( {\begin{smallmatrix}
		a_1  \\
		\vdots \\
		a_{|\MM|} \\
\end{smallmatrix} } \right)
$ is the exogenous-arrival-rate vector, $\bm{\lambda} = \left( {\begin{smallmatrix}
		\lambda_1  \\
		\vdots \\
		\lambda_{|M|} \\
\end{smallmatrix} } \right)$ is the demand rate vector, $\bm{I}$ is an identity matrix, and $\bm{R}$ is a $|\MM|\times|\MM|$ turning ratio matrix, whose locations of non-zero values are decided by the network structure. Note that for any traffic network, if the average exogenous arrival rates are all zeros ($\bm{a} = \bm{0}$), the average demand rates will also be zeros ($\bm{\lambda} = \bm{0}$). Hence, $(\bm{I} - \bm{R})$ must be invertible in traffic network, and we have $\bm{\lambda} = (\bm{I} - \bm{R})^{-1}\cdot\bm{a}$.

\begin{definition}[Stability Region]
	The network's stability region, denoted by $\mathcal{D}$, is defined as the 
	closed convex hull of the set of all demand rate vectors $\bm{\lambda} \in \RR_{\geq 0}^{|\MM|}$ for which there exists a stabilizing control policy under given technical background.
	\label{def2}
\end{definition}

The ``technical background'' refers to all the hardware and software technical requirements for implementation. For example, real-time control policies are infeasible if the controller cannot adjust the signal schemes dynamically, and predicted I-SFR is unavailable if the controller does not have an I-SFR prediction module. This paper focuses on investigating how the knowledge level of I-SFR would influence stability region and leaves the discussion on other techniques to the future.

\section{Changes in stability region}
\label{S:SR}
\subsection{I-SFR and M-SFR}
When a real-time controller is making decisions, it cares about the SFR of each movement if it were to receive a green signal immediately. In this context, the real-time SFR in the near future is defined as the \textit{imminent SFR (I-SFR)}. The I-SFR is a non-negative bounded iid random variable with a mean value equal to M-SFR. Its value is influenced by real-time traffic conditions. For each movement $m$, the set of all its possible I-SFR values is denoted by $S_m$. For a movement group $\MM$, we can express the set of all its possible joint I-SFR values (denoted by $E$) by $E = \bigtimes_{m \in \MM} S_m$. Taking $|E|$ as the total number of all possible joint values in $E$, we can calculate $m$'s M-SFR by 
\begin{equation} 
	\overline{s}_m = \EE \{s_m\} = \sum_{e=1}^{|E|} p^e s_m^e,
\end{equation}
where $s_m^e$ is $m$'s I-SFR in the $e$th joint value, and $p^e$ is the probability of the $e$th joint value of $E$ with $\sum_{e=1}^{|E|} p^e = 1$.

\subsection{$\mathcal{D}_0$: stability region with knowledge of M-SFR}
\label{SS:D0}
When the controller only knows M-SFR, the capacity of a movement can be obtained by timing its M-SFR and (effective) green ratio \citep{hcm2010}:
\begin{equation} 
	c_m = \overline{s}_m g_m,
	\label{E:capacity_1}
\end{equation}
where $c_m$ is $m$'s capacity considering M-SFR, and $g_m \in [0,1]$ is $m$'s 
green ratio. $g_m$ can be calculated as the expectation of $\phi_m(t)$: $g_m = \EE \{ \phi_m(t)\}$. Considering all movements within the network, we denote the green ratio vector as $\bm{g} = \left( {\begin{smallmatrix}
		g_1  \\
		\vdots \\
		g_{|\MM|} \\
\end{smallmatrix} } \right) $, which consists of all movements' green ratios. We denote a complete set of all possible $\bm{g}$ as $\GX$.
Generally, $\GX$ can be formulated as a linear constraint of $\bm{g}$:
\begin{equation} 
	\GX = \{\bm{g}: \bm{K} \cdot \bm{g} \leq \bm{h} \},
\end{equation}
where $\bm{K}$ is a constraint matrix for all conflicting movements and non-negative boundaries, and $\bm{h}$ is a (column) vector with proper dimensions. For example, for a single intersection with two conflicting through movements, we have $g_1 + g_2 \leq 1$, $g_1 \geq 0$, and $g_2 \geq 0$. Hence $\bm{K} =
\left( {\begin{smallmatrix}
		1 & 1 \\
		-1 & 0 \\
		0 &-1 \\
\end{smallmatrix} } \right)$ 
and $\bm{h} =
\left( {\begin{smallmatrix}
		1  \\
		0 \\
		0 \\
\end{smallmatrix} } \right)$. It is easy to verify that $\GX$ is a convex polyhedron region. 

According to \autoref{thm:1}, to stabilize the queues, the arrival rate of movement $m$ (denoted by $\lambda_m \geq 0$) should not exceed its capacity: $\lambda_m \leq c_m$. Therefore, the stability region only \ul{with the knowledge of M-SFR}, denoted by $\mathcal{D}_0$, can be expressed as 
\begin{equation} 
	\mathcal{D}_0 = \{(\bm{a},\bm{R}): (\bm{I} - \bm{R})^{-1}\cdot\bm{a} = \bm{\overline{s}}\cdot \bm{g},~~ \bm{g} \in \GX\},
	\label{E:D_0}
\end{equation}
where  $\bm{\overline{s}} = 
\left( {\begin{smallmatrix}
		\overline{s}_1 & &   \\
		& \ddots & \\
		& & \overline{s}_{|\MM|} \\
\end{smallmatrix} } \right) $ is the M-SFR matrix. 
Another popular expression for stability region is $\{(\bm{a},\bm{R}): \bm{0} \leq (\bm{I} -\bm{R})^{-1}\cdot\bm{a} \leq \overline{\bm{s}}\cdot \bm{g},~~ \bm{g} \in \GX\}$. 
\begin{proposition}
	Region $\{(\bm{a},\bm{R}): \bm{0} \leq (\bm{I} - \bm{R})^{-1}\cdot\bm{a} \leq \bm{\overline{s}}\cdot \bm{g},~~ \bm{g} \in \GX\}$ and region $\{(\bm{a},\bm{R}): (\bm{I} - \bm{R})^{-1}\cdot\bm{a} = \bm{\overline{s}}\cdot \bm{g},~~ \bm{g} \in \GX\}$  are the same.
	\label{P:D0}
\end{proposition}
The proof of \autoref{P:D0} is in \ref{Ap:P2}.

\subsection{$\mathcal{D}_1$: stability region with full knowledge of I-SFR}
\label{SS:D1}
When the controller has full knowledge of I-SFR, it can allocate different green ratios for different I-SFRs, hence
\begin{equation} 
	c_m = \EE\{s_m(t)\phi_m(t)\} = \sum_{e=1}^{|E|} p^e s_m^e g_m^e,
	\label{E:capacity_2}
\end{equation}
where $c_m$ is $m$'s capacity considering I-SFR, and $g_m^e \in [0,1]$ is $m$'s green ratio when the $e$th joint value happens: $g^e_m = \EE \{\phi_m(t) | e\}$. Clearly, $c_m$ in \eqref{E:capacity_1} and \eqref{E:capacity_2} are the same only when $s_m(t)$ and $\phi_m(t)$ are independent, which is true in traditional traffic scenarios. That is because the controller can hardly observe or estimate I-SFR with traditional technology. Hence it cannot adjust the green ratio according to I-SFR. 

Similarly, the stability region \ul{with full knowledge of I-SFR}, denoted by $\mathcal{D}_1$, can be expressed as 
\begin{equation} 
	\mathcal{D}_1 = \{(\bm{a},\bm{R}): (\bm{I} - \bm{R})^{-1}\cdot\bm{a} = \sum_{e=1}^{|E|} p^e\cdot\bm{s}^e\cdot \bm{g}^e,~~ \bm{g}^e \in \GX\},
	\label{E:D_1}
\end{equation}
where $\bm{s}^e = 
\left( {\begin{smallmatrix}
		s^e_1 & &   \\
		& \ddots & \\
		& & s^e_{|\MM|} \\
\end{smallmatrix} } \right) $ is the I-SFR matrix under the $e$th joint value. Note that \eqref{E:D_1} can also be expressed as $\{(\bm{a},\bm{R}): \bm{0} \leq (\bm{I} - \bm{R})^{-1}\cdot\bm{a} \leq \sum_{e=1}^{|E|} p^e\cdot\bm{s}^e\cdot \bm{g}^e,~~ \bm{g}^e \in \GX\}$.
We shall describe the relationship between two stability regions starting from defining the dominating point.
\begin{definition}[dominating point]
	A point $\alpha=(\alpha_1,\alpha_2,\cdots,\alpha_Z) \in \mathbb{R}^Z$ dominates a point $\beta=(\beta_1,\beta_2,\cdots,\beta_Z) \in \mathbb{R}^Z$, written as $\alpha \vdash \beta$ , if and only if for $z = 1, 2, \cdots, Z$, we have (a) $\forall z,~ \alpha_z \geq \beta_z$, and (b) $\exists z,~ \alpha_z > \beta_z$.
\end{definition}
Furthermore, similar to the definition of the Pareto frontier, we define the precise upper frontier of any region $\mathcal{A}$ in this paper.
\begin{definition} [Precise Upper Frontier]
	The precise upper frontier of region $\mathcal{A}$, denoted by $\bar{f}(\mathcal{A})$, is defined as:
	\begin{equation}
		\bar{f}(\mathcal{A}) = \big\{ \alpha' \in \mathcal{A}: \{\alpha'' \in \mathcal{A}: \alpha'' \vdash \alpha'
		\} = \emptyset  \big\}.
	\end{equation}
	\label{D:frontier}
\end{definition}			
The precise upper frontier of a stability region indicates the maximum supply ability of the network given a certain technical background. It is a bit different from the general upper frontier, whose concept will be elaborated on in \autoref{D:n-frontier} in the subsequent section.

\begin{example}
	\label{Ex:1}
Here we show an example about difference between $\mathcal{D}_0$ and $\mathcal{D}_1$ through a single intersection with only two conflicting through movements (hence $\bm{R}$ is a zero matrix, and we can just ignore it in the analysis). Suppose for each time interval, movement 1 has two possible I-SFR values, $1$ and $2$ (veh/time-interval) with a probability of 0.3 and 0.7, respectively; movement 2 has two possible I-SFR values, $1$ and $2$ with a probability of 0.5 and 0.5, respectively. Obviously, $|E| = 4$. We let $ (s^1_1,s^2_1,s^3_1,s^4_1)= (1,2,1,2)$ and $ (s^1_2,s^2_2,s^3_2,s^4_2)= (1,1,2,2)$, hence $(p^1, p^2, p^3, p^4) = ( 0.15, 0.35, 0.15, 0.35)$. The M-SFR matrix $\overline{\bm{s}} = \sum_{e=1}^4 p^e \left( {\begin{smallmatrix}
		s^e_1 &    \\
		&s^e_2    \\
\end{smallmatrix} } \right) = 
\left( {\begin{smallmatrix}
		\overline{s}_1^{} &    \\
		&\overline{s}_2^{}    \\
\end{smallmatrix} } \right) 
= \left( {\begin{smallmatrix}
		1.7_{}&    \\
		&1.5_{}    \\
\end{smallmatrix} } \right)$. We show $\mathcal{D}_0$ and $\mathcal{D}_1$ in \autoref{F:CR}. 
It is clear that,  $\mathcal{D}_0\subset\mathcal{D}_1$, and $\mathcal{D}_1$ has a larger upper frontier and hence a larger size (increased by $22.8\%$) compared with $\mathcal{D}_0$. Note that, with full knowledge of I-SFR, if we 
only assign green to movement 2 when movement 1's I-SFR is 1 and movement 2's I-SFR is 2, that is, we let $(\bm{g}^1,\bm{g}^2,\bm{g}^3,\bm{g}^4) = \left( {\begin{smallmatrix}
		1 & 1 & 0 & 1 \\
		0 & 0 & 1 & 0 \\
\end{smallmatrix} } \right)$, then capacity $\bm{c} = \left( {\begin{smallmatrix}
		0.7*2+0.15*1 \\
		0.15*2 \\
\end{smallmatrix} } \right) = \left( {\begin{smallmatrix}
		1.55 \\
		0.3 \\
\end{smallmatrix} } \right) $; if we 
only assign green to movement 1 when movement 2's I-SFR is 1 and movement 1's I-SFR is 2, that is, we let $(\bm{g}^1,\bm{g}^2,\bm{g}^3,\bm{g}^4) = \left( {\begin{smallmatrix}
		0 & 1 & 0 & 0 \\
		1 & 0 & 1 & 1 \\
\end{smallmatrix} } \right)$, then capacity $\bm{c} = \left( {\begin{smallmatrix}
		0.35*2 \\
		0.15*1+0.5*2 \\
\end{smallmatrix} } \right) = \left( {\begin{smallmatrix}
		0.7 \\
		1.15 \\
\end{smallmatrix} } \right) $.
\begin{figure}[h!]
	\centering
	\resizebox{0.6\textwidth}{!}{
		\includegraphics{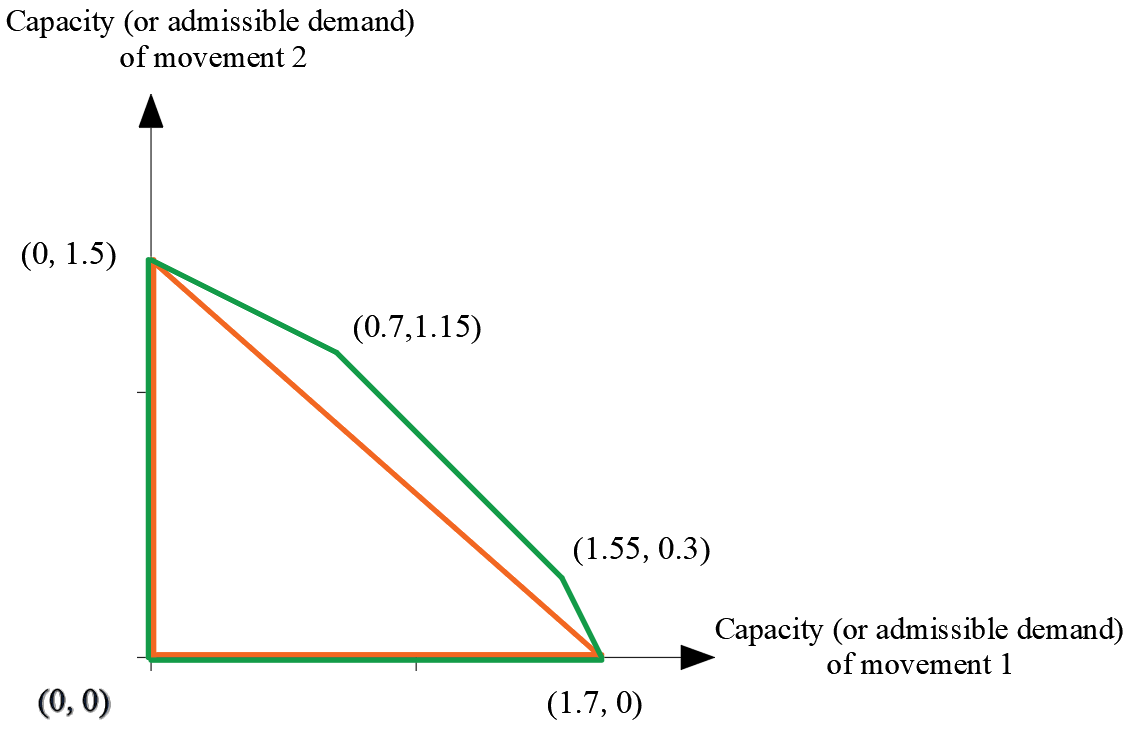}}
	\caption{Stability regions, $\mathcal{D}_0$'s hull: \deeporange{orange}, $\mathcal{D}_1$'s hull: \deepgreen{green}.} 
	\label{F:CR}
\end{figure}
\end{example}

		
\begin{lemma} 
	If SFR varies, we have $\mathcal{D}_0 \subset \mathcal{D}_1$, and there exist point $\alpha \in \bar{f}(\mathcal{D}_1)$ and point $\beta \in \bar{f}(\mathcal{D}_0)$ such that $ \alpha \vdash \beta $.
	\label{thm:2}
\end{lemma}	
	We shall prove this lemma through two steps: 1) prove at least $\mathcal{D}_0 \subseteq \mathcal{D}_1 $, 2) prove there exists a point in $\mathcal{D}_1$ that dominates a point in $\bar{f}(\mathcal{D}_0)$, and hence $\mathcal{D}_0 \subset \mathcal{D}_1 $.
	
\begin{proof}
	\textit{Step 1.} Because $\GX$ is a convex polyhedron region, given any $g \in \GX$, 
	we can always find $|E|$ number of $\bm{g}^e \in\GX$ to express $\bm{g}$ with
	\begin{equation} 
		\bm{g} = \sum_{e=1}^{|E|} p^e \cdot \bm{g}^e.
		\label{E:overline_s}
	\end{equation}
	Recall that $p^e$ is the joint value's probability with $\sum_e p^e = 1$, and we can let $ \bm{g}^1 = \cdots = \bm{g}^{|E|} =\bm{g}$ satisfy \eqref{E:overline_s}, then $\sum_{e=1}^{|E|} p^e\cdot\bm{s}^e\cdot \bm{g}^e = (\sum_{e=1}^{|E|} p^e\cdot\bm{s}^e)\cdot \bm{g} = \overline{\bm{s}} \cdot \bm{g}$. It means all points in $\mathcal{D}_0$ are also in $\mathcal{D}_1$. Hence, $\mathcal{D}_0 \subseteq \mathcal{D}_1$.
	
	\textit{Step 2.} We choose a $\bm{g}$ satisfying 1) $\bm{g} \in \bar{f}(\GX)$, and 2) $g_m<1$ for all $m$ owning conflicting movements. 
	From the nature of processing network, we know that increasing some turning ratios or/and exogenous demands (without decreasing the others) will always lead to the increase of arrivals in some movements, which is directly bounded by the movement's capacities. Hence $(\bm{a},\bm{R}):(\bm{I} - \bm{R})^{-1}\cdot\bm{a} = \overline{\bm{s}} \cdot \bm{g}$ with $\bm{g}\in \bar{f}(\GX)$ is at $\bar{f}(\mathcal{D}_0)$.
	We let $ \bm{g}^1 = \cdots = \bm{g}^{|E|} = \bm{g}$, hence $\sum_{e=1}^{|E|} p^e\cdot\bm{s}^e\cdot \bm{g}^e = \overline{\bm{s}} \cdot \bm{g}$. We then choose a group of collaborative movements $\{m_1, m_2, \cdots\}$ that can be discharged simultaneously in one interval, and choose another group of collaborative movements $\{\mu_1, \mu_2, \cdots\}$ so that $\{m_1, m_2, \cdots\}$ and $\{\mu_1, \mu_2, \cdots\}$ are conflicting. Denoting $\Delta g$ as a small positive value, we choose a joint value $e_1$ and add $\frac{\Delta g}{p^{e_1}}$ to $g^{e_1}_{m_1}$, $g^{e_1}_{m_2}$, $\cdots$ (the green ratio of movements $\{m_1, m_2, \cdots\}$ in joint value $e_1$) with an upper bound of one. Because of the constraint of conflicting movements, we need to subtract $\frac{\Delta g}{p^{e_1}}$ from $g^{e_1}_{\mu_1}, g^{e_1}_{\mu_2}, \cdots$ with a lower bound of zero. On the other hand, we choose another joint value $e_2$ and add $\frac{\Delta g}{p^{e_2}}$ to $g^{e_2}_{\mu_1}, g^{e_2}_{\mu_2}, \cdots$. Similarly, we need to subtract $\frac{\Delta g}{p^{e_2}}$ from  $g^{e_2}_{m_1}, g^{e_2}_{m_2}, \cdots$. The chosen $e_1$ and $e_2$ satisfy some conditions, as shown below: 
	\begin{equation*}
		\small
		{\begin{matrix}
				\\
				\\
				\cdots ~ \\
		\end{matrix} } {\begin{matrix}
				\mathrm{joint~value}~e_1\mathrm{:}~~ \\
				~\\
				\left( {\begin{matrix}
						\vdots\\
						g^{e_1}_{m_1} \\
						g^{e_1}_{m_2} \\
						\vdots \\
						g^{e_1}_{\mu_1}\\
						g^{e_1}_{\mu_2}\\
						\vdots \\
				\end{matrix} } \right) 
				{\begin{matrix}
						\\
						\green{+ \frac{\Delta g}{p^{e_1}}}\\
						\green{+ \frac{\Delta g}{p^{e_1}}}\\
						\\
						\orange{- \frac{\Delta g}{p^{e_1}}}\\
						\orange{- \frac{\Delta g}{p^{e_1}}}\\
						\\
				\end{matrix} }\\
		\end{matrix} } ~ {\begin{matrix}
				\\
				\\
				~~\cdots ~ \\
		\end{matrix} } {\begin{matrix}
				\mathrm{joint~value}~e_2\mathrm{:}~~ \\
				~\\
				\left( {\begin{matrix}
						\vdots\\
						g^{e_2}_{m_1} \\
						g^{e_2}_{m_2} \\
						\vdots \\
						g^{e_2}_{\mu_1}\\
						g^{e_2}_{\mu_2}\\
						\vdots \\
				\end{matrix} } \right) 
				{\begin{matrix}
						\\
						\orange{- \frac{\Delta g}{p^{e_2}}}\\
						\orange{- \frac{\Delta g}{p^{e_2}}}\\
						\\
						\green{+ \frac{\Delta g}{p^{e_2}}}\\
						\green{+ \frac{\Delta g}{p^{e_2}}}\\
						\\
				\end{matrix} }\\ 
		\end{matrix} }
		~~~ {\begin{matrix}
				\\
				\\
				~\cdots~, \text{~where the chosen $e_1$ and $e_2$ satisfy:} \\
		\end{matrix} }
		{\begin{matrix}
				~ \\
				~\\
				~	\\
				\left\{{\begin{matrix}
						s^{e_1}_{m_1} > s^{e_2}_{m_1} \\
						s^{e_1}_{m_2} > s^{e_2}_{m_2} \\
						\vdots \\
						s^{e_1}_{\mu_1} < s^{e_2}_{\mu_1}\\
						s^{e_1}_{\mu_2} < s^{e_2}_{\mu_2}\\
						\vdots \\ 
				\end{matrix} } \right. 
				\\ 
		\end{matrix} } {\begin{matrix}
				\\
				\\
				~. \\
		\end{matrix} }
	\end{equation*}
	Because I-SFRs of different movements with no shared lanes are independent with each other, we can at least find two joint values that satisfy the above inequality group. After the above modifications to green ratios of $e_1$ and $e_2$, we can calculate the new movement capacity for any $m \in \{m_1, m_2, \cdots\} \cup \{\mu_1, \mu_2, \cdots\}$ as:
	\begin{equation*}
		\small
		\left\{ {\begin{matrix}
				\underset{e}{\sum} p^e s^e_{m_1} g^{e,\mathrm{new}}_{m_1}=& \underset{e}{\sum} p^e s^e_{m_1} g^e_{m_1} + p^{e_1} s^{e_1}_{m_1} \frac{\Delta g}{p^{e_1}} - p^{e_2} s^{e_2}_{m_1} \frac{\Delta g}{p^{e_2}} =  \underset{e}{\sum} p^e s^e_{m_1} g^e_{m_1}  +\Delta g(s^{e_1}_{m_1} -s^{e_2}_{m_1}) > \underset{e}{\sum} p^e s^e_{m_1} g^e_{m_1}\\
				\vdots&\\
				\underset{e}{\sum} p^e s^e_{\mu_1} g^{e,\mathrm{new}}_{\mu_1}=& \underset{e}{\sum} p^e s^e_{\mu_1} g^e_{\mu_1} - p^{e_1} s^{e_1}_{\mu_1} \frac{\Delta g}{p^{e_1}} + p^{e_2} s^{e_2}_{\mu_1} \frac{\Delta g}{p^{e_2}} =  \underset{e}{\sum} p^e s^e_{\mu_1} g^e_{\mu_1}  +\Delta g(s^{e_2}_{\mu_1} -s^{e_1}_{\mu_1}) > \underset{e}{\sum} p^e s^e_{\mu_1} g^e_{\mu_1}\\
				\vdots&\\
		\end{matrix} }\right. . 
	\end{equation*} 
	
	As we can see, the capacity of each movement $m \in \{m_1, m_2, \cdots\} \cup \{\mu_1, \mu_2, \cdots\}$ is further increased as a result of the modifications. 
	This means a new capacity dominating the old one is achieved by adjusting the green ratios in different joint values with knowledge of I-SFR, and hence we can easily find a point in $\bar{f}(\mathcal{D}_1)$ dominating a point in $\bar{f}(\mathcal{D}_0)$, indicating that $\mathcal{D}_0 \subset \mathcal{D}_1 $. 
\end{proof}

In general, \autoref{thm:2} implies that $\mathcal{D}_1$ has a higher upper frontier than $\mathcal{D}_0$. Hence, if a technology can help the controller obtain accurate I-SFR, the stability region of the network can be enlarged compared with only using M-SFR. 
\begin{example}
    Here we show a numerical example for Step 2 of \autoref{thm:2}'s proof 
    to find $\bar{f}(\mathcal{D}_0)$'s dominating point in $\mathcal{D}_1$.	
	
	Continue with the settings in \autoref{Ex:1}. 
	We choose an initial point at precise upper frontier of $\mathcal{D}_0$ with $\bm{g}^e = \bm{g} = \left( \begin{smallmatrix} 0.7\\
		0.3\\ 
	\end{smallmatrix}\right)$, and the corresponding capacity vector $\bm{c} = \left( {\begin{smallmatrix}
			0.7*1.7 \\
			0.3*1.5 \\
	\end{smallmatrix} } \right) = \left( {\begin{smallmatrix}
			1.19 \\
			0.45 \\
	\end{smallmatrix} } \right)$. Take movement 1 as $m_1$ and movement 2 as $\mu_1$ in the proof, and take joint value 2 as $e_1$ and joint value 3 as $e_2$. Moreover, we set $\Delta g = 0.105$. After the modifications to $\bm{g}^2$ and $\bm{g}^3$, we have $(\bm{g}^1,\bm{g}^{2,\mathrm{new}},\bm{g}^{3,\mathrm{new}},\bm{g}^4) = \left( {\begin{smallmatrix}
			0.7 & 1 & 0 & 0.7 \\
			0.3 & 0 & 1 & 0.3 \\
	\end{smallmatrix} } \right)$. The new capacity vector $\bm{c}^{\mathrm{new}} = \left( {\begin{smallmatrix}
			0.15*1*0.7 + 0.35*2*1 + 0 + 0.35*2*0.7 \\
			0.15*1*0.3 + 0 + 0.15*2*1 + 0.35*2*0.3 \\
	\end{smallmatrix} } \right) = \left( {\begin{smallmatrix}
			1.295 \\
			0.555 \\
	\end{smallmatrix} } \right) >\bm{c} $. \autoref{F:CR_move} shows that $\bm{c}^{\mathrm{new}}$ moves to $\mathcal{D}_1$'s precise upper frontier coincidentally.		
	\begin{figure}[h!]
		\centering
		\resizebox{0.61\textwidth}{!}{
			\includegraphics{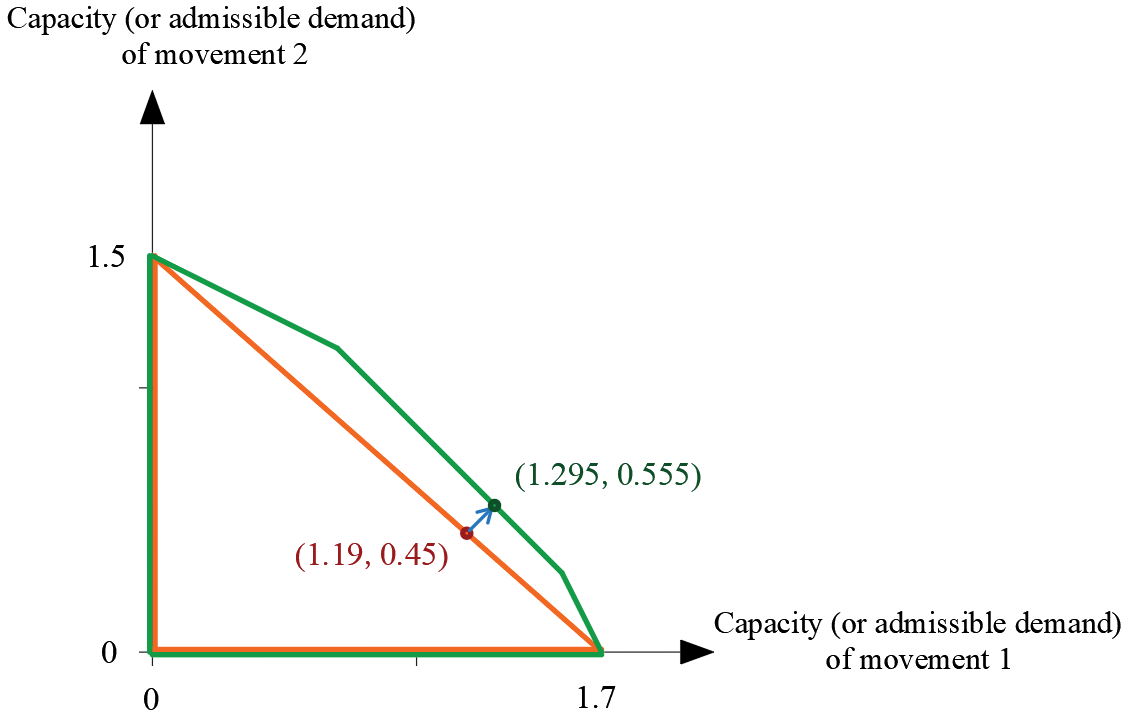}}
		\caption{Change from $\bm{c}$ to $\bm{c}^{\mathrm{new}}$ in Example 2.} 
		\label{F:CR_move}
	\end{figure}
\end{example}

\subsection{$\mathcal{D}_\theta$: stability region with partial knowledge of I-SFR}			
As stated above, even with advanced technologies, it is still hard to acquire full knowledge of I-SFR. Usually, the I-SFR can only be successfully predicted with a certain probability. 
We suppose there is a coefficient $\theta \in [0,1]$ to describe \emph{the ability of technology to ``forecast" I-SFR}. $\theta = 0$ means the technology has no ability to predict I-SFR; $\theta = 1$ means the technology can always know the I-SFR. If  $0<\theta<1$, the technology has a certain ability to predict I-SFR but may fail with a probability of $1-\theta$. In this paper, we assume that the prediction is unbiased. Clearly, a technology with larger $\theta$ indicates its prediction ability is stronger. If the I-SFR is in the state of ``unknown'' or ``fail of prediction'', the controller will still pick or guess a SFR value (e.g., the M-SFR) to replace the I-SFR. For a movement $m$, we denote $\hat{p}({\hat{s}_m})$ as the probability measure of guessing $\hat{s}_m$, and $g(\hat{s}_m)$ as the corresponding green ratio in guess, where $\int_{\hat{s}_m} \dd \hat{p}(\hat{s}_m)= 1$, and the green ratio strategy in randomly guess is $g_m=\int_{\hat{s_m}} g(\hat{s}_m) \dd\hat{p}(\hat{s}_m)$. For example, the simplest unbiased strategy is guessing the I-SFR equal to M-SFR ($\overline{s}_m$), i.e., $\hat{p}(\overline{s}_m) = 1$; another simple unbiased strategy is to make the guessing probability coincident with the real probability in historical data, $\hat{p}(s_m) = p(s_m)$. Note that the guess should also be unbiased.

In an extreme case with $\theta=0$, the controllers always need to ``guess'', and I-SFR information is not used. Its capacity vector $\bm{c}$ is
\begin{equation} 
	\bm{c}|_{\theta=0} = \underset{\text{$e$ happens~}}{\underbrace{\sum_{e=1}^{|E|}p^{e} \cdot \bm{s}^{e} \cdot}} \underset{\text{guess I-SFR}}{\underbrace{\int_{\underset{}{\hat{\bm{s}}}} \bm{g}(\hat{\bm{s}})\dd\hat{p}(\hat{\bm{s}})}} = \overline{\bm{s}} \cdot \bm{g}, ~~~~\bm{g}^e, ~ \bm{g} \in\GX, 
	\label{E:theta_0}
\end{equation}
which means that such a random guess when $\theta = 0$ is equivalent to the condition with only knowledge of M-SFR in \eqref{E:capacity_1} for each movement. Given $0<\theta<1$, the capacity vector $\bm{c}$ is
\begin{multline} 
	\bm{c}|_{\theta} = \EE\{\bm{s}(t)\bm{\phi}(t|\theta)\} = \sum_{e=1}^{|E|}p^{e} \cdot \bm{s}^{e} \cdot \big[\theta \cdot \bm{g}^{e} + (1-\theta) \cdot \bm{g} \big] \\= \theta \cdot \sum_{e=1}^{|E|}p^e \cdot \bm{s}^e \cdot \bm{g}^e  + (1-\theta) \cdot \overline{\bm{s}} \cdot 
	\bm{g}, ~~~~ \bm{g}^e, \bm{g} \in\GX, 
	\label{E:theta}
\end{multline}
where $\bm{\phi}(t)= \left( {\begin{smallmatrix}
		\phi_1(t)  \\
		\vdots \\
		\phi_{|\MM|}(t) \\
\end{smallmatrix} } \right)$. 
Because the implementation of any control strategy is limited by the level of prediction technique (represented by $\theta$), we express the control strategy $\bm{\phi}(t|\theta)$ by $\bm{\phi}(t)$ for simplification in the remaining of this paper.

When the controller can predict the I-SFR with an ability coefficient of $0\leq\theta\leq1$, the corresponding stability region of the network, denoted by $\mathcal{D}_\theta$, can be expressed by
\begin{equation} 
	\mathcal{D}_\theta = \big \{(\bm{a},\bm{R}): (\bm{I}-\bm{R})^{-1}\bm{a} = \theta \cdot \sum_{e=1}^{|E|}p^e \cdot \bm{s}^e \cdot \bm{g}^e  + (1-\theta) \cdot \overline{\bm{s}} \cdot
	\bm{g},~~ \bm{g}^e, \bm{g} \in \GX\big|\theta \in [0,1]\big\}.
	\label{E:D_theta}
\end{equation}
Note that $\mathcal{D}_\theta$ with $\theta=0$ or $\theta=1$ is just equivalent to the $\mathcal{D}_0$ or $\mathcal{D}_1$ discussed in \autoref{SS:D0} and \autoref{SS:D1}. 

\begin{example}
	Here we show an example for $\mathcal{D}_\theta$. 
	Continued with the settings in 
	\autoref{Ex:1} (and \autoref{F:CR}), \autoref{F:CR_half_theta} additionally shows the stability region when $\theta=0.5$. It is clear that, the precise upper frontier of $\mathcal{D}_\theta$ is between the precise upper frontiers of $\mathcal{D}_0$ and $\mathcal{D}_1$, and its size is increased by 11.2\% compared with $\mathcal{D}_0$.
	
	\begin{figure}[h!]
		\centering
		\resizebox{0.61\textwidth}{!}{
			\includegraphics{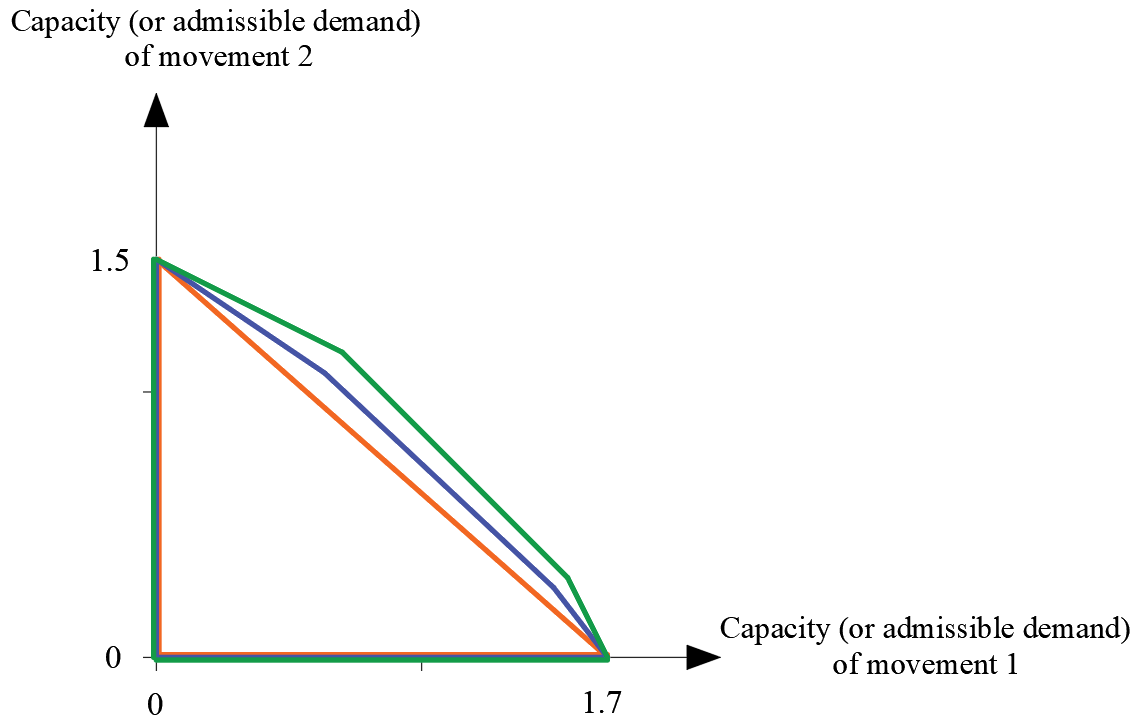}}
		\caption{Stability region $\mathcal{D}_\theta$'s hull with $\theta = 0$: \deeporange{orange}, $\theta = 0.5$: \deepblue{blue}, and $\theta = 1$: \deepgreen{green}.}
		\label{F:CR_half_theta}
	\end{figure}
\end{example}	

\begin{theorem}
	For any real numbers $\theta_\mathrm{a}$ and $\theta_\mathrm{b}$ with $0 \leq \theta_\mathrm{a} < \theta_\mathrm{b} \leq 1$, we have $\mathcal{D}_{\theta_\mathrm{a}} \subset \mathcal{D}_{\theta_\mathrm{b}}$, and there exist points $\alpha_{\mathrm{a}}\in \bar{f}(\mathcal{D}_{\theta_\mathrm{a}})$ and $\alpha_{\mathrm{b}}\in \bar{f}(\mathcal{D}_{\theta_\mathrm{b}})$ such that $\alpha_{\mathrm{b}} \vdash \alpha_{\mathrm{a}}$.
	\label{C:D_theta}
\end{theorem}			
\begin{proof}
	Based on \autoref{thm:2}, we can always find a green ratio strategy with knowledge of I-SFR to let $\sum_{e=1}^{|E|} p^e\cdot\bm{s}^e\cdot \bm{g}^e \geq \overline{\bm{s}} \cdot \bm{g}$ be true, and there always exists $\sum_{e=1}^{|E|} p_m^e s_m^e g_m^e -\overline {s}_m\cdot g_m > 0$ when $\bm{g}$ is not at a vertex of $\GX$. Therefore, given $\theta_\mathrm{a}$ and $\theta_\mathrm{b}$ with $\forall 0 \leq \theta_\mathrm{a} < \theta_\mathrm{b} \leq 1$, we can always find a green ratio strategy with knowledge of I-SFR to let $\bm{c}|_{\theta_\mathrm{b}} ~-~ \bm{c}|_{\theta_\mathrm{a}} = (\theta_\mathrm{b}-\theta_\mathrm{a})(\sum_{e=1}^{|E|} p^e\cdot\bm{s}^e\cdot \bm{g}^e - \overline{\bm{s}} \cdot \bm{g}) \geq \bm{0}$ be true, where ``$>$'' holds true for some movements. Since their lower bounds are the same, we have $\mathcal{D}_{\theta_\mathrm{a}} \subset \mathcal{D}_{\theta_\mathrm{b}}$, and $\mathcal{D}_{\theta_\mathrm{b}}$ has a higher upper frontier than $\mathcal{D}_{\theta_\mathrm{a}}$, i.e., $\exists\alpha_{\mathrm{b}} \in \bar{f}(\mathcal{D}_{\theta_\mathrm{b}}) \vdash \alpha_{\mathrm{a}} \in \bar{f}(\mathcal{D}_{\theta_\mathrm{a}})$.
\end{proof}

One may note that the prediction ability $\theta$ is not straightforward because we can only obtain the \emph{prediction accuracy} from the data in practice. Even if a technology has no ability to predict, the accuracy of prediction could still be larger than 0 with an appropriate guess strategy. Therefore, it is worth analyzing the relationship between the prediction ability and the prediction accuracy.
\begin{proposition}
	The prediction ability and prediction accuracy are positively correlated.
	\label{prop: pred}
\end{proposition}

The proof of \autoref{prop: pred} is in \ref{Ap:P3}. \autoref{prop: pred} implies that improving the prediction accuracy actually improves the prediction ability and hence enlarges the precise upper frontier of the stability region.

\section{Network control and stability}
\label{S:BP}

\subsection{SFR-only algorithm and reserve demand}
We build a network-wide SFR information-only (SFR-only) policy, which only uses the predicted I-SFR among all real-time information. It can stabilize the queueing network whenever the demands are inside the stability region. Based on \eqref{E:D_theta}, when the predicted I-SFR is obtained given a prediction ability of $\theta$, the SFR-only algorithm seeks to maximize $\epsilon$ with constraints to ensure that the demand is within the stability region:
\begin{align}
	&&\text{Max:}&&&&  	&&\epsilon& \\
	&&\text{s.t.:}&&&& (\bm{I} - \bm{R})^{-1}&\cdot(\bm{a} +\bm{\epsilon}) &\leq& ~\theta \cdot \sum_{e=1}^{|E|}p^e \cdot \bm{s}^e \cdot \bm{g}^e  + (1-\theta) \cdot \overline{\bm{s}} \cdot \bm{g},&&&&&&&&  \label{E:constraint1}\\
	&&&&&& &\bm{K} \cdot \bm{g}^e &\leq& ~\bm{h},~ e = \emptyset,  1,2,\cdots, |E|.& &&&&&&&&
\end{align}
The known parameters that appear as constants in the above linear program include the exogenous-arrival-rate vector $\bm{a}$, the turning ratio matrix $\bm{R}$, the prediction ability $\theta$, the joint value probability $p^e, e = 1, 2, \cdots, |E|$, the I-SFR matrix $\bm{s}^e, e = 1, 2, \cdots, |E|$, the M-SFR matrix $\overline{\bm{s}}$, and the conflicting movement matrix/vector $\bm{K}$/$\bm{h}$. The unknowns that act as variables to be optimized in the above linear program include: the green ratio vector $\bm{g}^e, e = \emptyset, 1, 2, \cdots, |E|$, and 
$\bm{\epsilon} = \left( {\begin{smallmatrix}
		\epsilon  \\
		\vdots \\
		\epsilon \\
\end{smallmatrix} } \right)$ with $|\MM|$ rows.

Define $\epsilon_{max}$ as the maximum value of $\epsilon$ in the above problem. The value of $\epsilon_{max}$ represents a measure of the distance between the demand rate vector $\bm{\lambda} = (\bm{I} - \bm{R})^{-1}\cdot\bm{a}$ and the general upper frontier (defined as below) of the stability region ($\mathcal{D}_{\theta}$), where $\bm{\lambda} \geq \bm{0}$. We name $\epsilon_{max}$ as the \emph{reserve demand} under the given demand rate. 
\begin{definition} [General Upper Frontier]
	\label{D:n-frontier}
	The general upper frontier of a closed region $\AX$, denoted by $f(\AX)$, is defined as:
	\begin{equation}
		f(\AX) = \big\{ \alpha' \in \AX: \{\alpha'': \alpha' \leq \alpha'' \leq \alpha' + \bm{\epsilon}\} \cap \AX^{\mathrm{C}} \ne \emptyset, ~\forall \epsilon>0  \big\},
	\end{equation}
	where $\AX^{\mathrm{C}}$ is complementary set
	of $\AX$.
\end{definition}			

\begin{proposition}
	$\bar{f}(\AX) \subseteq f(\AX).$
\end{proposition}
\begin{proof}
	Suppose $\alpha \in \AX \backslash f(\AX)$, from \autoref{D:n-frontier} we know, $\exists \epsilon>0$ and $\beta \in \{\alpha \leq \beta \leq \alpha+ \bm{\epsilon} \} \backslash \alpha$ to let $\beta \in \AX$. As $\beta \vdash \alpha$, we know $\alpha \not \in \bar{f}(\AX)$. Hence, $\bar{f}(\AX) \subseteq f(\AX)$.
\end{proof}

Since $\bar{f}(\AX) \subseteq f(\AX)$, \underline{raising the precise upper frontier also raises the general upper frontier.} Intuitively speaking, the precise upper frontier of the stability region implies that \underline{all} movements are operating at their maximum supply abilities, while the general upper frontier of the stability region suggests that \underline{at least one} movement is operating at its maximum supply ability.

Let $\mathcal{D}_{\theta}^-$ denote the region $\mathcal{D}_{\theta}$ excluding its general upper frontier: $\mathcal{D}_{\theta}^- \equiv \mathcal{D}_{\theta}\backslash f(\mathcal{D}_{\theta})$. It is worth mentioning that $\mathcal{D}_{\theta}^-$ differs slightly from the interior of  $\mathcal{D}_{\theta}$ as $\mathcal{D}_{\theta}^-$ includes the lower bound, such as the point of admissible demand ``zero'', whereas the interior of $\mathcal{D}_{\theta}$ excludes them. If $\bm{\lambda}$ lies in $\mathcal{D}_{\theta}^-$, then $\epsilon_{max} > 0$ and the network is stable; if $\bm{\lambda}$ locates at $f(\mathcal{D}_{\theta})$, then $\epsilon_{max} = 0$ and the network is also stable; if $\bm{\lambda}$ (non-negative) is outside of $\mathcal{D}_{\theta}$, then $\epsilon_{max} < 0$ and the network can not be stable. Clearly, when $\epsilon_{max} \geq 0$, there always exists a SFR-only phasing scheme $\Phi^{\mathrm{SFR-only}}$ (and corresponding green ratios $\bm{g}^e$) that makes the network rate stable. We denote $\bm{\phi}^{\mathrm{SFR-only}}(t)$ as the vector of all movements' passing status when $\Phi^{\mathrm{SFR-only}}$ is implemented by the controller.

Note that although the SFR-only algorithm is stabilizing, it can hardly be used in the real world since it requires prior knowledge of future arrival rate and the entire probability distribution of the network's joint SFR for all time. This is unrealistic in practice since traffic operation environment can change hour by hour. Therefore, we only regard the SFR-only algorithm as a "springboard" to prove BP's stability. It also helps to show the relationship between the stability and reserve demand. 

\begin{example}
	Here we show an example of the SFR-only algorithm about how the reserve demand changes with $\theta$. 
	Consider a simple network with only two nodes (intersections) and eight movements as shown in \autoref{F:Network_simple}. The exogenous-arrival-rate vector $\bm{a} = (2,1,0,0,1.6,1,0,0)^\mathrm{T}$, and the turning ratio vector for movements $\{1, 2, \cdots, 8\}$ is $(r_1,r_2,\cdots,r_8) = (0,0,0.25,0.75,0,0,0.2,0.8)$. Considering the network structure, the turning ratio matrix can be written as
	$\bm{R} = \left( { \begin{smallmatrix}
			0 & 0 & 0 & 0 & 0 & 0 & 0 & 0 \\
			0 & 0 & 0 & 0 & 0 & 0 & 0 & 0 \\
			0 & 0 & 0 & 0 &r_3& 0 &r_3& 0 \\ 
			0 & 0 & 0 & 0 &r_4& 0 &r_4& 0 \\
			0 & 0 & 0 & 0 & 0 & 0 & 0 & 0 \\
			0 & 0 & 0 & 0 & 0 & 0 & 0 & 0 \\
			r_7& 0 &r_7& 0 & 0 & 0 & 0 & 0 \\ 
			r_8& 0 &r_8& 0 & 0 & 0 & 0 & 0 \\  
	\end{smallmatrix} } \right)$. Assume the controller makes decisions every 10 s, and for every movement $m$, we suppose its I-SFR every 10 s can be 3 or 4 (veh/10 s) with an equal probability of 0.5. 
	
	Each node has three possible phases, excluding movement groups with merging conflicts, as shown in \autoref{F:Network_simple}. We have $g_1^e+g_3^e \leq 1$, $g_1^e+g_4^e \leq 1$, $g_2^e+g_4^e\leq 1$ for Node 1, and $g_5^e+g_7^e \leq 1$, $g_5^e+g_8^e \leq 1$, $g_6^e+g_8^e\leq 1$ for Node 2. The conflicting matrix $ \bm{K} = \left( {\begin{matrix}
			\bm{K}'\\
			-\bm{I} \\
	\end{matrix} } \right)$, where $\bm{K}' = \left( 
	{ \begin{smallmatrix}
			1 & 0 & 1 & 0 & 0 & 0 & 0 & 0 \\
			1 & 0 & 0 & 1 & 0 & 0 & 0 & 0 \\
			0 & 1 & 0 & 1 & 0 & 0 & 0 & 0 \\
			0 & 0 & 0 & 0 & 1 & 0 & 1 & 0 \\
			0 & 0 & 0 & 0 & 1 & 0 & 0 & 1 \\
			0 & 0 & 0 & 0 & 0 & 1 & 0 & 1 \\
	\end{smallmatrix} } \right)$, and the conflicting vector is $\bm{h} = (1,1,1,1,1,1,0,0,0,0,0,0,0,0)^\mathrm{T}$.
	\begin{figure}[h!]
		\centering
		\resizebox{0.7\textwidth}{!}{
			\includegraphics{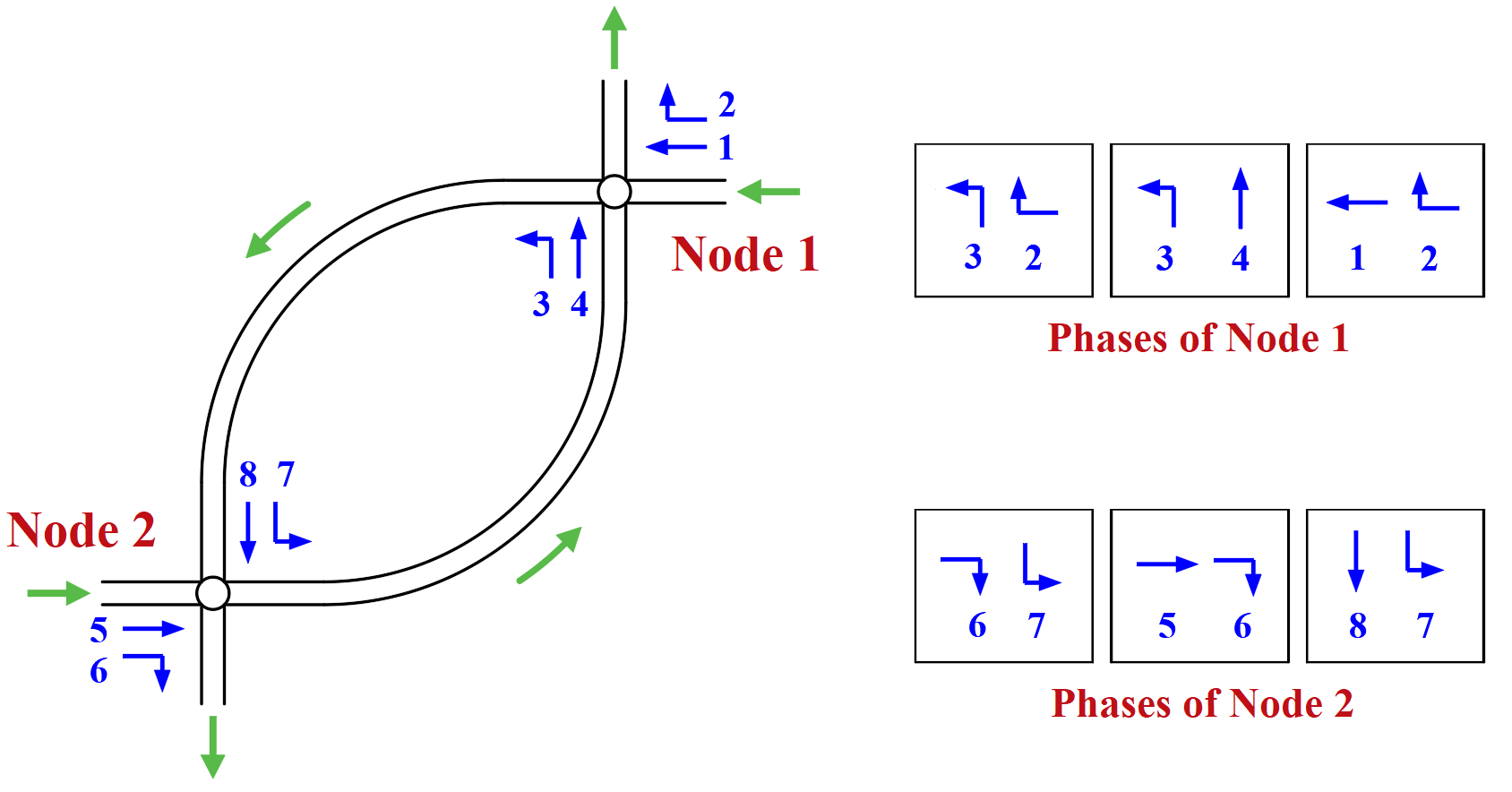}}
		\caption{Structure of a simple network and its possible phases in Example 4.}
		\label{F:Network_simple}
	\end{figure}
	
	When $\theta$ takes different values from 0 to 1, $\epsilon_{max}$ of this network will also be different. We draw $\epsilon_{max}$ with different $\theta$ values in \autoref{F:ex4_theta}. Obviously, $\epsilon_{max}$ increases linearly with $\theta$, and it equals to 0 when $\theta = 0.485$. Since the network can only be stabilized (for all movements) when $\epsilon_{max} \geq 0$, we have that the network can be rate table only when $\theta \geq 0.485$. When $\theta < 0.485$, there is always an unstable movement (at least) for all possible effective green ratios, and hence no control algorithm can stabilize the network.
	\begin{figure}[h!]
		\centering
		\resizebox{0.55\textwidth}{!}{
			\includegraphics{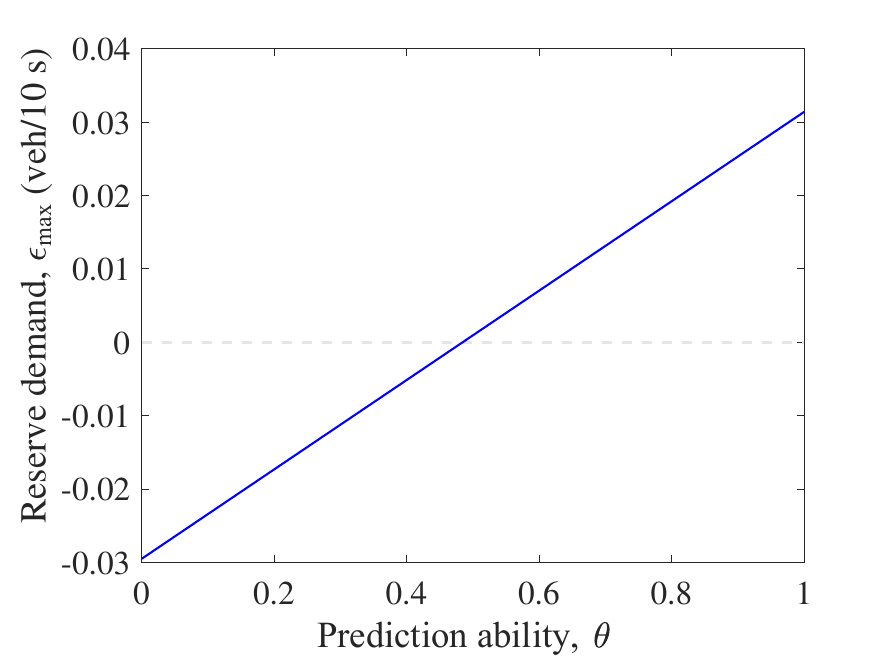}}
		\caption{Changes of $\epsilon_{\max}$ with different $\theta$ in Example 4.}
		\label{F:ex4_theta}
	\end{figure}
\end{example}

\begin{lemma}
	Assume that the demand rate lies in $\mathcal{D}_{\theta}^-$, and the SFR-only strategy, $\Phi^{\mathrm{SFR-only}}$, is adopted. Then, there exists a constant $\epsilon > 0$ such that $\EE \{ (\bm{I}-\bm{R}(t)) \cdot \bm{s}(t) \cdot \bm{\phi}^{\mathrm{SFR-only}}(t) - \bm{a}(t) \} \geq \bm{\epsilon}$ holds. 
	\label{L:SFR-only}
\end{lemma}

The proof of \autoref{L:SFR-only} is in \ref{Ap:L2}.

\subsection{BP control with predicted I-SFR}
At the beginning of each time interval, BP controllers make decisions based on the current queueing state and the predicted I-SFR of this interval. Original BP control assumes that controllers can predict I-SFR with an accuracy of 100\% (e.g., \cite{wongpiromsarn2012distributed}). Let $\Phi^{\mathrm{BP,1}}_n(t) \in \bm{\Phi}_n$ denote the phasing scheme of intersection $n$ chosen by BP with accurate I-SFR at time $t$.  It is the phasing scheme that solves the following problem:
\begin{equation}
	\Phi^{\mathrm{BP,1}}_n(t) \in \underset{\Phi_n \in \bm{\Phi}_n}{\arg \max} \sum_{m\in \MM_n} w_m^{\mathrm{BP}}(t) s_m(t)\phi_m(t), \label{E:BP}
\end{equation}
where $s_m(t)$ is the real I-SFR of movement $m$ during interval $[t,t+1)$, and $w_m^{\mathrm{BP}}(t)$ is the \textit{weight variable} defined by
\begin{equation}
	w_m^{\mathrm{BP}}(t) \equiv  x_m(t) -  \sum_{j \in O(m)} x_j(t) r_j(t). \label{E:BPwt}
\end{equation}
Obviously, BP is a distributed policy allowing intersections to make decisions independently based on neighboring information. Since the number of possible phases at any intersection tends to be small (typically four-eight, see \autoref{F:intersection_example}\subref{F:II}), \eqref{E:BP} can be easily solved by direct enumeration. Though simple, BP with accurate I-SFR comes with robust theoretical guarantee, which says as long as the demand vector belongs to $\mathcal{D}^-_1$, BP with accurate I-SFR can ensure strong stability of the network.
\begin{definition}[Strong stability]
	The traffic network is said to be \textit{strongly stable}:
	\begin{equation}
		\underset{T \rightarrow \infty}{\lim \sup} ~ \frac{1}{T} \sum_{t=0}^{T-1} \sum_{m \in \MM} \EE \{x_m(t)\}< \infty. \label{E:stable}
	\end{equation}
\end{definition}~

We refer the readers to \cite{wongpiromsarn2012distributed} for the stability proof of BP with accurate I-SFR. Clearly, strong stability is stronger than rate stability, and it implies rate stability. The strong stability in the proof requires $\epsilon>0$, which is stricter than the requirement of the rate stability ($\epsilon\geq0$).

As stated above, logically $s_m(t)$ cannot be directly observed in advance, and the controller can only predict I-SFR with an ability of $\theta$. We propose to choose the phase by solving the following problem:
\begin{equation}
	\Phi^{\mathrm{BP},\theta}_n(t) \in \underset{\Phi_n \in \bm{\Phi}_n}{\arg \max} \sum_{m\in \MM_n} w_m^{\mathrm{BP}}(t) \hat{s}_m(t)\phi_m(t), \label{E:BP_theta}
\end{equation}
where $\hat{s}_m(t)$ is the predicted I-SFR of movement $m$ during interval $[t,t+1)$, and $w_m^{\mathrm{BP}}(t)$ is the weight variable in \eqref{E:BPwt}.
We define the error item, $\mE$, as
	\begin{equation}
		\mE_m(t) \equiv \hat{s}_m(t) - s_m(t).
	\end{equation} 
From the unbiasedness of $\hat{s}_m(t)$ under green, we have $\EE \{\mE_m(t)| \phi_m(t)=1\} = 0$.
	\begin{lemma}
		\label{L:Unbias}
		If the predicted I-SFR is unbiased, there is $\EE \big\{\hat{s}_m(t)\phi_m(t) - a_m(t) - r_m(t) \sum_{i \in U(m)}  \hat{s}_i(t)\phi_i(t)  \big| \bm{X}(t)\big\} = \EE \big\{ s_m(t)\phi_m(t) -a_m(t)-r_m(t)\sum_{i \in U(m)}  s_i(t)\phi_i(t)  \big| \bm{X}(t)\big\}$. 
	\end{lemma}
The proof of \autoref{L:Unbias} is in \ref{Ap:L3}.

We define the network-wide Lyapunov functional ($L$) as
\begin{equation}
	L \big( \boldsymbol{X}(t) \big) \equiv  \frac{1}{2} \sum_{m \in \MM} x_m^2(t), \label{E:Lyapunov}
\end{equation}
and define the Lyapunov drift ($\Delta$) as
\begin{equation}
	\Delta \big( \boldsymbol{X}(t) \big) \equiv  \EE \{ L \big( \boldsymbol{X}(t+1) \big) -  L\big( \boldsymbol{X}(t) \big) | \boldsymbol{X}(t) \}. \label{E:Drift}
\end{equation}
\autoref{thm:LyapunovDrift} below provides a sufficient condition for strong stability using the definition of Lyapunov functionals. 

\begin{lemma}
	\label{thm:LyapunovDrift}
	For the Lyapunov functional \eqref{E:Lyapunov}, suppose $\EE \{L \big( \boldsymbol{X}(0) \big) \} < \infty$.  If there exist constants $0 < K < \infty$ and $0 < \epsilon <\infty$ such that
	\begin{equation}
		\Delta \big( \boldsymbol{X}(t) \big) \le K - \epsilon \sum_{m \in \MM} x_m(t) \label{E:stCond}
	\end{equation}
	holds for all $t \ge 0$ and all possible $\boldsymbol{X}(t)$, then the traffic network is strongly stable.
\end{lemma}
The proof of \autoref{thm:LyapunovDrift} is in \ref{Ap:L4}.
\begin{theorem}[Stability of BP with predicted I-SFR]
	\label{thm:BP_theta}
	Assume that arrival rates lie in $\mathcal{D}_\theta^-$, then the BP policy with unbiased I-SFR prediction \eqref{E:BP_theta} ensures strong stability of the traffic network.
\end{theorem}
The proof of \autoref{thm:BP_theta} uses \autoref{L:SFR-only}, \autoref{L:Unbias} and \autoref{thm:LyapunovDrift}. The proof details are in \ref{Ap:T2}. \autoref{thm:BP_theta} says that in spite of the changes in stability region with different prediction abilities, the BP can still maximize the network throughput. In other words, if the technique can further increase the I-SFR's prediction ability, the following enlarged stability region can be sensitively captured by the BP with predicted I-SFR. This interesting property can be revealed by the changed reserve demands in a field calibrated simulation, and BP policies in it serves as a probe to sense the changes in stability region. 

\section{Experiments}
\label{S:simulation}
This section shows through experiments how I-SFR knowledge can affect the stability region and the control effectiveness. We use three methods to predict the I-SFR, which produces significantly different accuracies. We test the performance of three control policies under the three accuracy levels in calibrated simulations, using the field fixed-time control as a comparison baseline. The analyzed index includes reserve demand and average vehicle delay.
\subsection{Simulation setting}
\begin{figure}[h!]
	\centering
	\resizebox{0.74\textwidth}{!}{
		\includegraphics{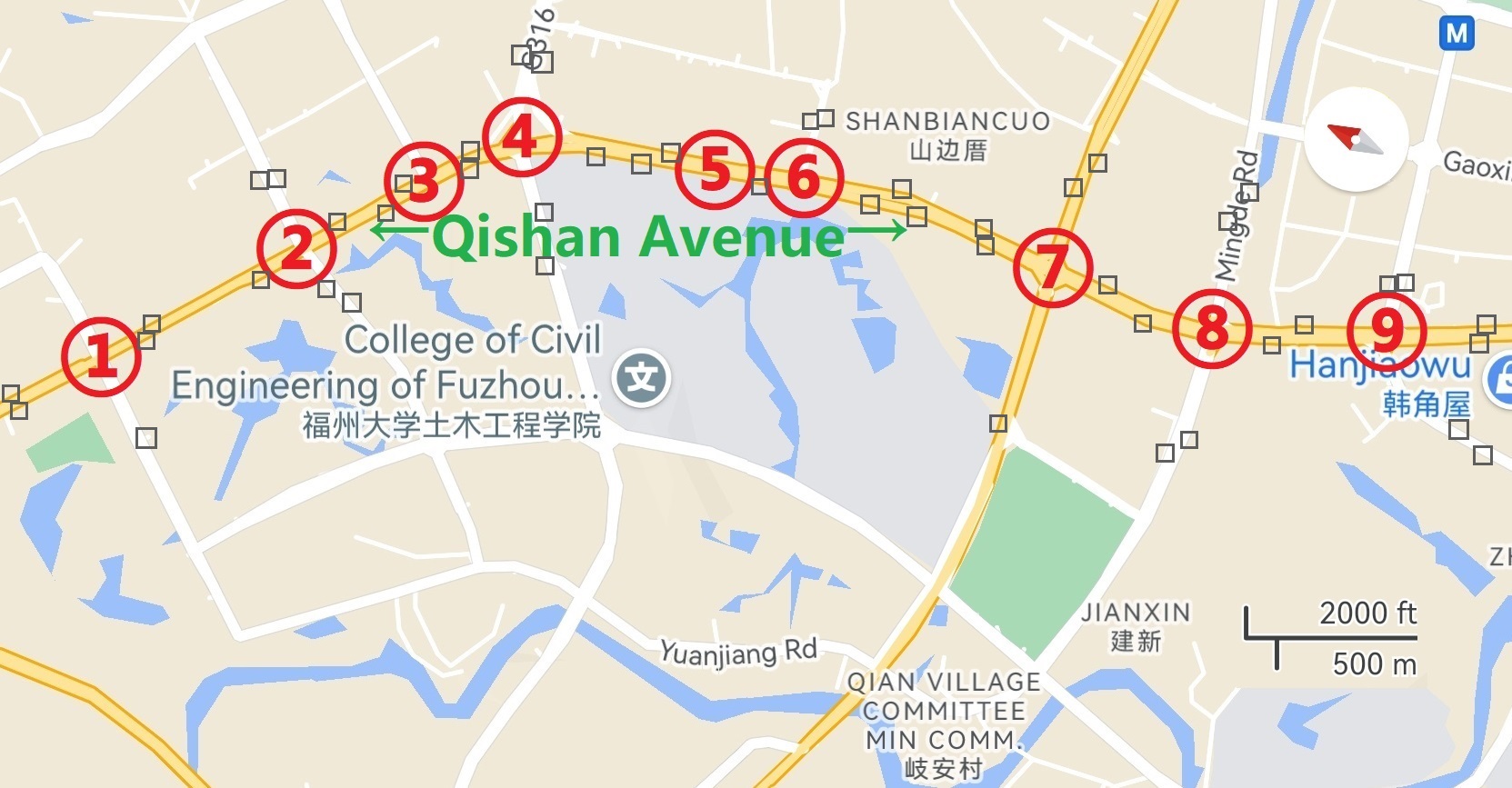}}
	\caption{Simulation network topology of Qishan Avenue in Fuzhou.
	} 
	\label{F:Network}
\end{figure}

\begin{figure}[h!]
	\centering
	\resizebox{0.75\textwidth}{!}{
		\includegraphics{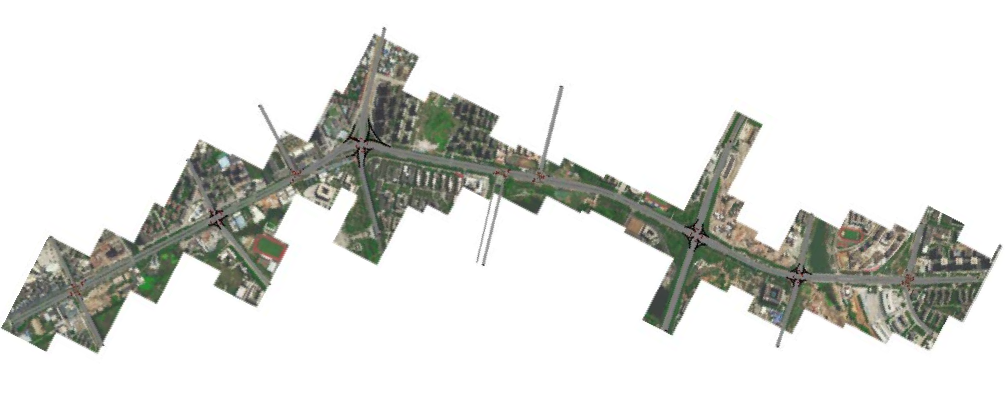}}
	\caption{Simulation model.} 
	\label{F:Vissim}
\end{figure}

\autoref{F:Network} shows the network used in our simulations, which is a corridor with nine intersections on Qichan Avenue, Fuzhou, Fujian, China. With the historical data collected during the evening peak hour (from 17:30 to 18:30) from Oct. 24 (Monday) to 28 (Friday), 2022, using videos, we calibrate the demands (including the vehicle compositions) of both ends of the avenue and all cross-streets. In additions, for different approaches of all nine intersections, we calibrate the turning ratios of all movements. The vehicular demands vary greatly from 18 veh/h/lane to 410 veh/h/lane, and the demand patterns are shown in \ref{Ap:demand}. The network topology, including the road lengths, widths, directions, lane layouts, and bus stops along the way, are extracted from the BaiduMap with the assistance of field surveys. The field controllers along Qishan Avenue use fixed-time control with signal coordination, and the cycle time during peak hour is as long as 196s. We built the simulation model with a microscopic simulation tool, VISSIM, as shown in \autoref{F:Vissim}. The field fixed-time control plan is reproduced in the simulation to serve as the benchmark. For BP-based policies, we set the time interval to 10 s, i.e., the BP controllers make phase decisions every 10 s. Using the COM interface of VISSIM, we feed the real-time collected data from VISSIM to Python and then return the optimized policy from Python to VISSIM.

In order to collect the I-SFR data when a movement is discharged, we only collect the samples when the queue length is greater than or equal to 7 vehs/lane (considering the time interval is 10 s, the max I-SFR per lane could be as large as 7 vehs/lane. If we collect samples with queue length $<7$ vehs/lanes, the samples and trained models will have bias).  
Considering the different volumes of non-motorized vehicles for each movement, we collect vehicle discharging data during the first 10 s of green and calibrate each movement's corresponding start-up lost time separately. The collected throughput data is then converted to I-SFR (veh/10-s/movement) of integers to make them consistent with the samples collected during the non-initial green phase (when the previous phase is the same as the current one). Note that for the non-initial green phase, the start-up lost time is 0 s.

We adopt three kinds of methods to predict the I-SFR, as listed below:
\begin{itemize}
	\item ``\underline{mean}'': always take the M-SFR as the predicted I-SFR;
	\item ``\underline{est}'': estimate the I-SFR as the weighted average of the last four historical I-SFRs;
	\item ``\underline{nn}'': learn the I-SFR using the deep neural network.
\end{itemize}

\begin{figure}[h!]
	\centering
	\resizebox{0.95\textwidth}{!}{
		\includegraphics{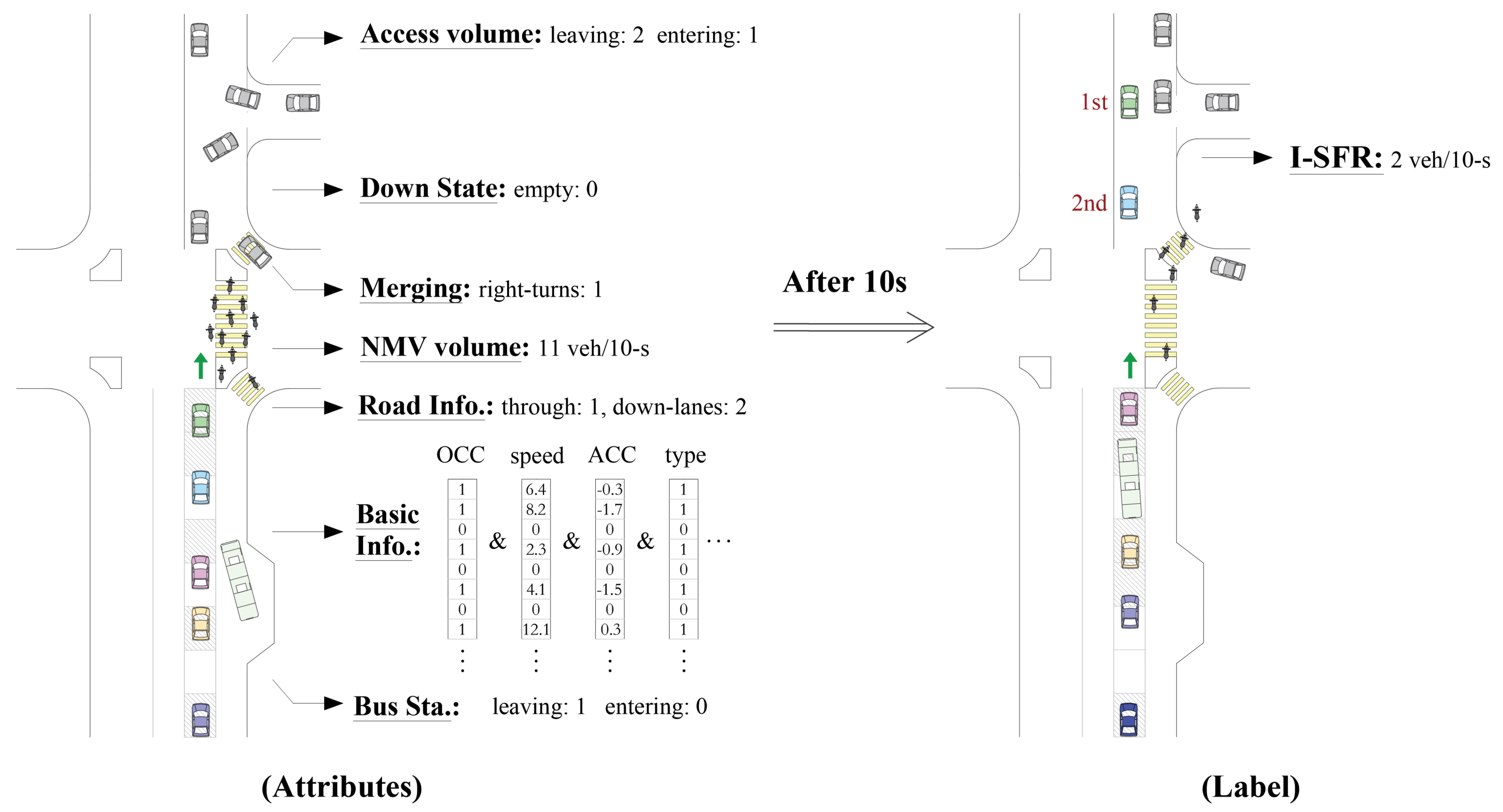}}
	\caption{Data extraction for training of the neural network model. 
	} 
	\label{F:influence}
\end{figure}

For the three prediction methods, we note that:
\begin{enumerate}[label=\arabic*)]
	\item When calculate the M-SFR, for each movement, we aggregate all I-SFR samples and take the average value of all I-SFRs as the M-SFR.
	\item When calibrating the weights for different historical I-SFRs in the "est" method, we test different weight combinations and choose [1, 2, 3, 4] as the final choice. This means the most recent I-SFR has the highest weight of 4, followed by the second closest I-SFR with a weight of 3, and so on.
	\item When learn the I-SFR using deep neural network, we collect different attributes, as shown in \autoref{F:influence}. We divide the road into cells of equal length (the average space that a vehicle occupies when traffic is in jam density), and extract the basic information for each cell, including occupancy (1 if occupied by a vehicle, 0 otherwise), vehicle speed (0 if empty), vehicle acceleration (0 if empty), and vehicle type. Besides, we also consider the non-motorized vehicle volume, the number of merging right-turn vehicles, the lane information, the number of bus entering or leaving the bus station (if any) within the 10 s free-flow distance, the number of vehicles within the 10 s free-flow distance of downstream link, and the number of vehicles entering or leaving the access (if any) within the 10 s free-flow distance. When training the machine learning model, the input of each sample is the attribute vector, and the label is the converted I-SFR per 10 s per movement.
	\item In order to extract I-SFR samples, we run the simulation for 10 times with different random seeds and each time lasts for 24 hours with different demand levels every 4 hours. The original BP with a predetermined M-SFR is used for generating the samples. We collect separate samples and calculate M-SFR for each movement. We also train separate learning models for each movement. We adopt the deep neural network with multi-classification as the learning model, which has one input layer, three hidden layers, and one output layer. The model is fully connected, and we pick the ReLU as its activation function in hidden layers. For the output layer, we use the softmax to activate it. The Adam optimizer is used with learning rate annealing. We use the batch normalization for each hidden layer to accelerate the training of neural network, and adopt the dropout technique for each training iteration to randomly drop neurons (with 50\% probability) from the neural network to prevent over-fitting. 80\% of the samples are used as the training set, and the rest 20\% are used as the testing set.
\end{enumerate}

\autoref{T:acc} shows the prediction accuracy and prediction time of different methods. The statistical results are based on different movements. Note that although the collected real I-SFRs are integers, non-integers are allowed in I-SFR's prediction since BP does not require the SFR to be integers. When calculating accuracy, we regard the prediction to be accurate when the absolute value of error is not greater than 0.5 veh/10-s/movement. As we can see, "mean" performs the worst with a lowest average accuracy of 13.1\%. With the help of more recent information, "est" improves the accuracy to 24.7\%. Moreover, with a more comprehensive consideration of different real-time factors, the machine learning technique "nn" succeeds in significantly improving the average accuracy to 56.7\%. Since the three methods have obvious difference in prediction accuracy, we are able to use them to test the influence of different levels of I-SFR knowledge on the network performance. 

It is worth mentioning that the goal of the experiments is not to design a perfect prediction model but to validate the proposed theory. To achieve the goal, we "intentionally" develop three prediction methods with significant differences in accuracy. In addition, although a more complicated prediction method tends to have a longer computation time, all three methods cost less than 0.01 s when running on a laptop equipped with a RTX2060 GPU processor and 12GB of memory. Hence they all have potential to be implemented in real time. We note that the "nn" method can compute quickly because the model is used online but trained offline. In practice, the training task can be assigned to the server. The controller only needs to update the trained model with low frequency (e.g., every hour), or the controller can train the model itself at midnight.
\begin{table}[!ht]
	\caption{Prediction accuracy of different methods.}
	\centering
	\setlength{\extrarowheight}{2pt}
	\begin{tabular}{c|ccccc|cc}
		\hline
		\multirow{2}{*}{Method:} & \multicolumn{5}{c|}{\underline{~Prediction accuracy:~}}  & \multicolumn{2}{c}{\underline{Prediction time (s):}}  \\ 
		& average & median & SD     & max   & min & average & SD \\ \hline
		``mean''  & 0.131   & 0.131  & 0.026 & 0.182 & 0.064  & 7.0E-07 & 1.7E-07\\
		``est''  & 0.247   & 0.235  & 0.057 & 0.441 & 0.179  & 3.9E-06 & 2.5E-07\\
		``nn''   & 0.567   & 0.567  & 0.073 & 0.800 & 0.465  & 7.5E-03 & 4.5E-05\\ \hline
	\end{tabular}
	\label{T:acc}
\end{table}

\subsection{Simulation controls}

We use the filed fixed-time control as the baseline, and test three BP control policies: the BP control in \eqref{E:BP_theta} and two other BP variants. 
\begin{itemize}
	\item \underline{``BP''}: original BP with predicted I-SFR, as shown in \eqref{E:BP_theta}.
	\item \underline{``PWBP''}: position-weighted BP with predicted I-SFR, in which vehicles closer to the controlled intersection will have higher weight. 
	\item \underline{``LESCBP''}: learning-extended switching-curve-based BP with predicted I-SFR, in which the controller under a higher volume will switch phase with a lower frequency to reduce the time loss.
\end{itemize}

PWBP assigns weights to vehicles based on their distance to the intersection so as to discharge more urgent vehicles \citep{li2019position} . Furthermore, it relaxes the M-SFR assumption, replaces the M-SFR by the actual number of passing vehicles in methodology, and designs a heuristic method to predict actual vehicles in simulation. However, the prediction of actual number of passing vehicles is more complicated than the prediction of I-SFR. Besides, it uses absolute value in weight calculation, which would increase the delay (without absolute value, its stability proof still works in \cite{li2019position}'s Equation $(58)$). Hence this paper in simulation slightly changed the format of PWBP's phase strategy by replacing actual passing vehicles with I-SFR and removing its absolute value:
\begin{equation}
	\Phi^{\mathrm{PWBP}}_n(t) \in \underset{\Phi_n \in \bm{\Phi}_n}{\arg \max} \sum_{m\in \MM_n} w_m^{\mathrm{PWBP}}(t) \hat{s}_m(t)\phi_m(t),
	\label{E:PWBP}
\end{equation}
with $w_m^{\mathrm{PWBP}}(t)$ defined by
\begin{equation}
	w_m^{\mathrm{PWBP}}(t) \equiv \sum_{v \in V_m(t)} \Big|\frac{d(v)}{l(m)}\Big|  -  \sum_{j \in O(m)} \sum_{v' \in V_j(t)} \Big|\frac{l(j) - d(v')}{l(j)}\Big| ~r_j(t),
	\label{E:PWBPwt}
\end{equation}
where $v$ or $v'$ is a vehicle, $V_m(t)$ is all vehicles queueing in $m$ at $t$ (hence $|V_m(t)| = x_m(t)$), $l(m)$ is length of the link that $m$ originates from, $d(v)$ is the distance from the most-upstream point of vehicle $v$'s located link to $v$'s current location. The motivation behind \eqref{E:PWBP}-\eqref{E:PWBPwt} is that vehicles near the intersection are more urgent and hence deserve higher weights. The rigorous stability proof of PWBP can be found in \cite{li2019position}.

LESCBP assumes that each time the signal switches the phase, it would experience a fixed switching loss during which no vehicles could pass the intersection \citep{wang2022learning}. Vehicles can discharge (with M-SFR) only after the switching loss. They introduced a switching-curve and formulated the switching-curve-based BP (SCBP) to prevent the signal from switching too frequently. The SCBP were then extended to policy-gradient reinforcement learning SCBP (Learned-Extended-SCBP, LESCBP) for practical implementation (BackPressure is also called Max Pressure in some literature, e.g., LESCBP is named as LESCMP (Learned-Extended-Switching-Curve based Max Pressure) in \cite{wang2022learning}, and this paper uses the name of BP uniformly for the sake of clarity). 
With LESCBP, each intersection can either choose a phase to maximize the pressure function or keep the phase strategy unchanged, and we further replace its M-SFR by predicted I-SFR: 
\begin{equation}
	\Phi_n^{\mathrm{LESCBP}}(t) \in 
	\begin{cases} 
		\underset{\Phi_n \in \bm{\Phi}_n}{\arg\max} \underset{m\in \MM_n}{\sum} w_m^{\mathrm{LESCBP}}(t) \hat{s}_m\phi_m(t), & \mbox{ if } \psi_n(t)>0 \\ 
		\Phi_n^{\mathrm{LESCBP}}(t-1), & \mbox{ if } \psi_n(t)\leq 0 
	\end{cases},
	\label{E:LESCBP}
\end{equation}
where $w_m^{\mathrm{LESCBP}}(t)$ is the weight coefficient with a slight difference from $w_m^{\mathrm{PWBP}}(t)$ in the position weights for stopped and moving vehicles, and $\psi_n(t)$ is the phase-switch indicator calculated by 
\begin{equation}
	\psi_n(t) = \max_{\Phi_n \in \bm{\Phi}_n} \sum_{m\in \MM_n} w_m^{\mathrm{LESCBP}}(t) \hat{s}_m\phi_m(t) - \Big(\sum_{m\in \MM_n} w_m^{\mathrm{LESCBP}}(t) \hat{s}_m\phi_m(t-1) +  \alpha \lVert \bm{x}_n(t) \rVert^{\beta} \Big), 
\end{equation}
where $\alpha \lVert \bm{x}_n(t) \rVert^{\beta}$ is the switching-curve function, $\alpha$ and $\beta$ are positive constants that need to be learned (using reinforcement learning) and calibrated ($\alpha=0.05$ and $\beta=0.1$ in our simulation), and
$\lVert \bm{x}_n(t) \rVert$ is 1-norm of the column vector that equals the summation of all the queue lengths of intersection $n$. With the switching rule given by \eqref{E:LESCBP}, the switching frequency will decrease with an increase in traffic demand, since a higher traffic demand usually leads to longer queue lengths, making the switching condition harder to be satisfied.

\subsection{Evaluation of reserve demand}
Our first experiment is to gauge the reserve demand ($\epsilon_{max}$) of the network, given the filed calibrated traffic demand. A higher $\epsilon_{max}$ implies a larger stability region boundary. 
In the simulation model, when the exogenous vehicles cannot enter the link due to congestion, they will temporally stack out of link. We assume that, once the demand is beyond the stability region, the stacked vehicle number will keep increasing. Therefore, we gradually increase the exogenous demand rate of each movement, and monitor the number of stacked vehicles in real time. Once the stacked vehicle number exceeds a preset threshold, we think that the demand has reached the stability region boundary. 

We set the initial demand to the real demand, and increase the exogenous demand rate by 5 vehs/h every 1 minute. The threshold is set to 100 vehs. That is, when the total stacked vehicles' number exceeds 100 vehs, we think the corresponding real-time demand reached the stability region boundary. The increased demand (the gap between the initial demand and the real-time demand) is regarded as $\epsilon_{max}$. More discussions about threshold choosing can be found in \ref{Ap:choice}.

\begin{figure}[!ht]
	\centering
	\subfloat[][BP]{\resizebox{0.3425\textwidth}{!}{
			\includegraphics[width=0.5\textwidth]{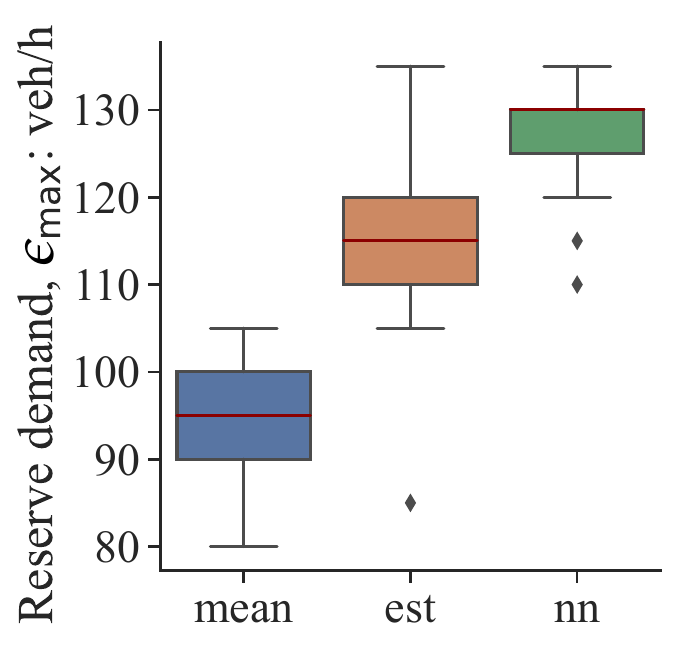}}
		\label{F:bp}} 
	~~
	\subfloat[][PWBP]{\resizebox{0.31\textwidth}{!}{
			\includegraphics[width=0.4\textwidth]{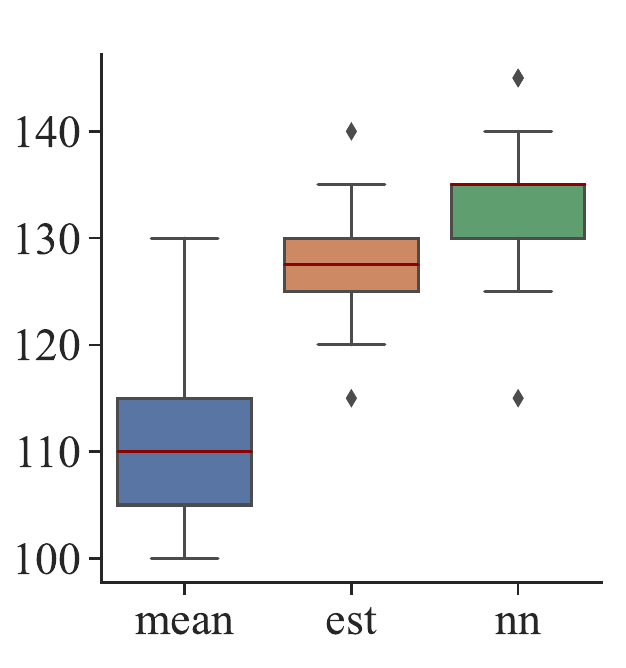}}
		\label{F:pwbp}}
	~~
	\subfloat[][LESCBP]{\resizebox{0.31\textwidth}{!}{
			\includegraphics[width=0.4\textwidth]{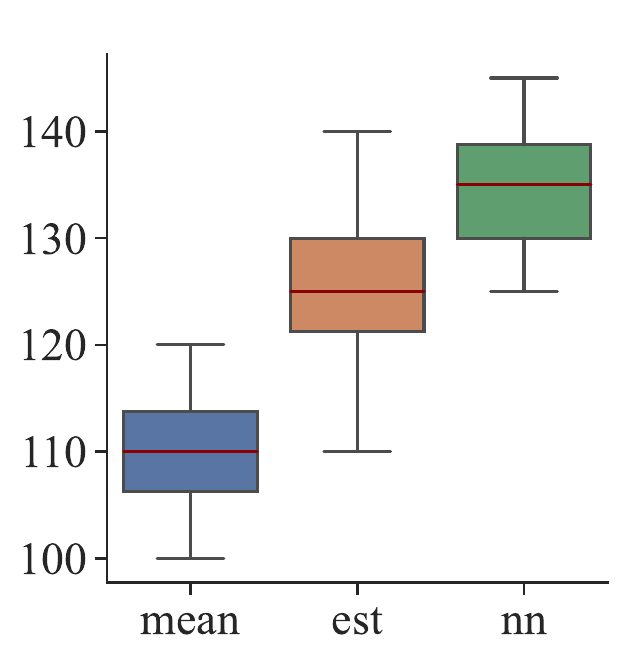}}
		\label{F:lescbp}}
	\caption{Reserve demand for three control policies with different I-SFR prediction methods.}
	\label{F:cr}
\end{figure}

For each combination of control policy and prediction method, we run the simulation for 30 times with different random seeds. 
The resulted $\epsilon_{max}$ are demonstrated with box plots, as shown in \autoref{F:cr}. As we can see, for all three BP-based policies, $\epsilon_{max}$ increases when the method can predict I-SFR more accurately. Take the original BP as an example: when the method changes from "mean" to "est" and "nn", $\epsilon_{max}$ is increased by about 21\% and 34\%, respectively, implying a continuous enlargement of the stability region. This experiment result is consistent with \autoref{C:D_theta}. We note that a more detailed inspection of the accuracy of different intersections shows that, for the critical intersection (the one that becomes congested earliest), the average accuracy of "est" is about 28\%, while the average accuracy of "nn" is just 47\%. This might be why $\epsilon_{max}$ increase from "est" to "nn" is not as noticeable as "expected".

\subsection{Impacts on average delay}
Our second experiment aims to investigate the delay performance with different I-SFR prediction accuracies. Delay is a very important index for evaluating the performance of a traffic network. However, we only proved the queueing stability of BP with predicted I-SFR, how the delay will change with the knowledge of I-SFR is still unknown. Therefore, we want to answer this question through experiments.

\begin{figure}[!ht]
	\centering
	\subfloat[][50\% real demand]{\resizebox{0.49\textwidth}{!}{
			\includegraphics[width=0.5\textwidth]{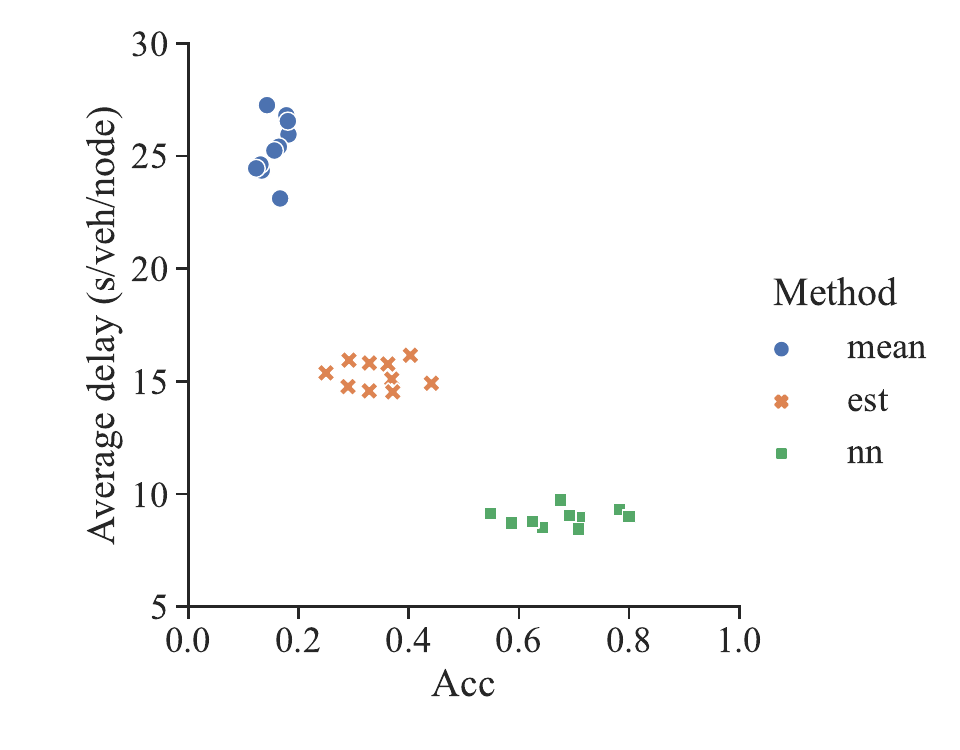}}
		\label{F:bp_delay0_5}} 
	~~
	\subfloat[][100\% real demand]{\resizebox{0.49\textwidth}{!}{
			\includegraphics[width=0.4\textwidth]{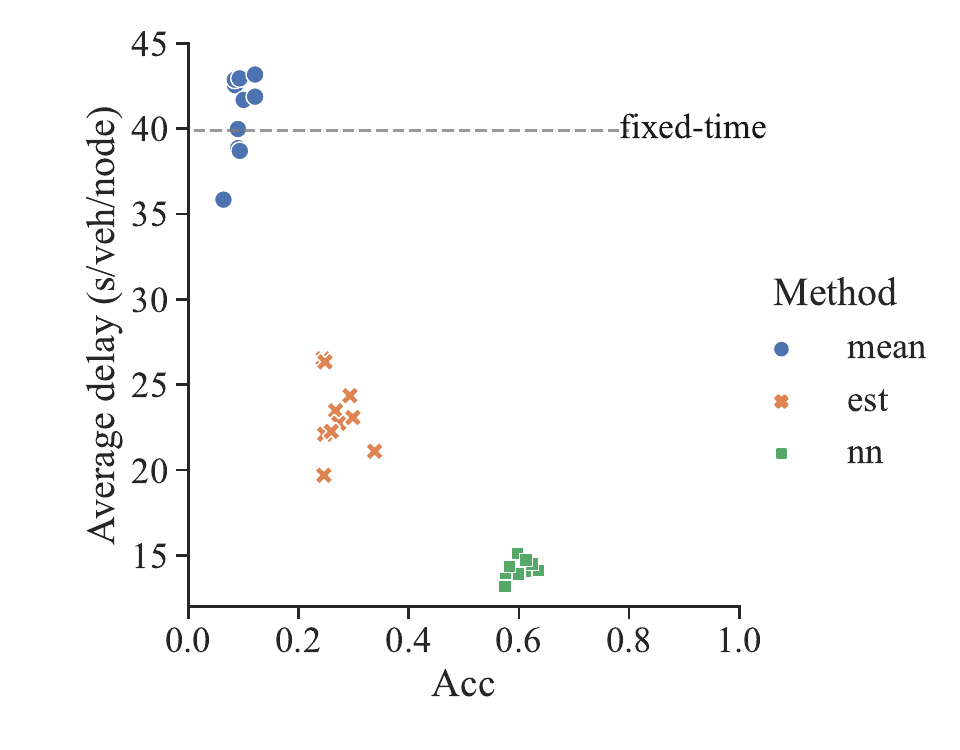}}
		\label{F:bp_delay1}}
	\caption{Delay and accuracy for BP control.}
	\label{F:BP}
\end{figure}

\autoref{F:BP} shows the average delay of BP with different I-SFR prediction methods when the demand is set to \subref{F:bp_delay0_5} 50\% and \subref{F:bp_delay1} 100\% levels of field demand. We note that the delay only considers motorized vehicles including cars and buses. Again, for each method, we run the simulation ten times with different random seeds, and each simulation lasts for one hour. The x-axis of \autoref{F:BP} represents the average prediction accuracy of different movements within the network, and the y-axis represents the average delay of all vehicles per intersection. For a better comparison, we show the average delay (around 40 s) resulting from the fixed-time control in the field with a dashed line when the demand is set to the field demand. As we can see, for both demands, the mean-BP (BP using a prediction method of ``mean'') has the highest delay. With a higher I-SFR prediction accuracy, est-BP reduces the delay to a much lower level. In addition, with a further improvement in prediction accuracy, nn-BP achieves the lowest delay. Obviously, more knowledge of I-SFR can effectively help BP reduce vehicle delay. Note that when the demand is half of the actual demand, the prediction accuracies of different methods are more sparse. This is because lower demand results in a smaller I-SFR sample size; hence the accuracy is less stable.

\begin{figure}[!ht]
	\centering
	\subfloat[][100\% real demand]{\resizebox{0.49\textwidth}{!}{
			\includegraphics[width=0.5\textwidth]{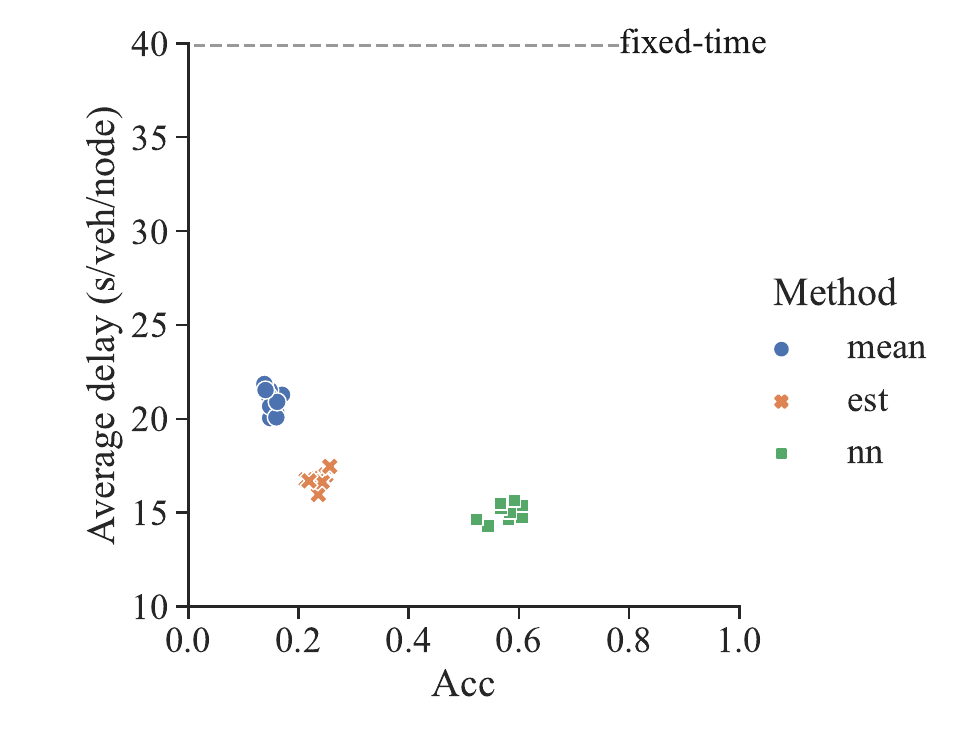}}
		\label{F:pwbp_delay1}} 
	~~
	\subfloat[][150\% real demand]{\resizebox{0.49\textwidth}{!}{
			\includegraphics[width=0.4\textwidth]{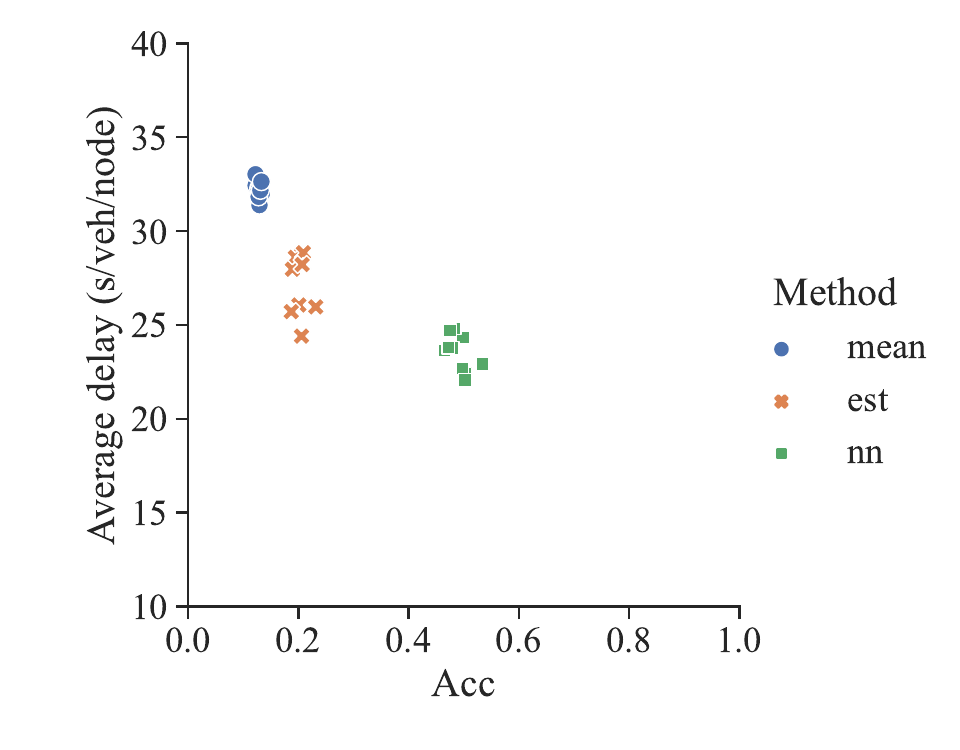}}
		\label{F:pwbp_delay1.5}}
	\caption{Delay and accuracy for PWBP control.}
	\label{F:pwbp_delay}
\end{figure}

\autoref{F:pwbp_delay} \subref{F:pwbp_delay1} shows the average delay of PWBP with different I-SFR prediction methods when the demand is set to the field demand. On one hand, we can notice that PWBP performs much better than BP in general, especially when using the methods ``mean'' and ``est''. This is consistent with the findings in \cite{li2019position}. On the other hand, we can also see a clear tendency that the delay of PWBP decreases with an increase in I-SFR prediction accuracy. Since the delay of mean-PWBP is just slightly higher than 20 s (much lower than the fixed-time delay), we further increase the demand input by 50\% to see the network performance of PWBP in more congested scenarios, as shown in \autoref{F:pwbp_delay} \subref{F:pwbp_delay1.5}. As we can see, the delay of PWBP with different methods increases as expected, while the tendency of delay with the change of accuracy remains unchanged. Hence we can conclude that more knowledge of I-SFR can help PWBP reduce vehicle delay.

\begin{figure}[!ht]
	\centering
	\subfloat[][100\% real demand]{\resizebox{0.49\textwidth}{!}{
			\includegraphics[width=0.5\textwidth]{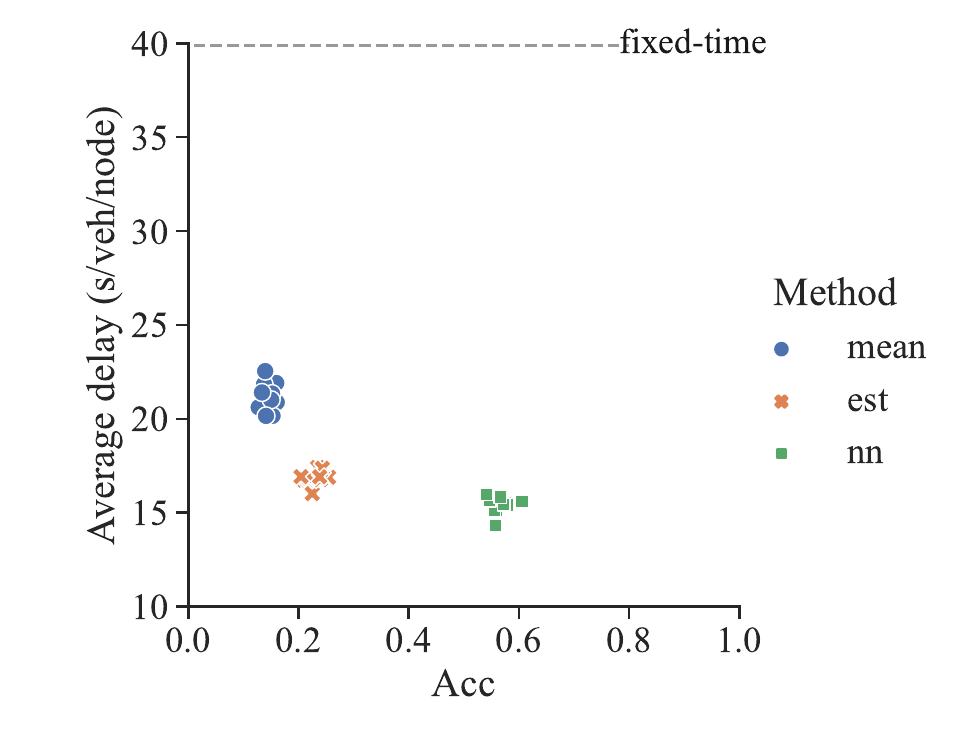}}
		\label{F:lescbp_delay1}} 
	~~
	\subfloat[][150\% real demand]{\resizebox{0.49\textwidth}{!}{
			\includegraphics[width=0.4\textwidth]{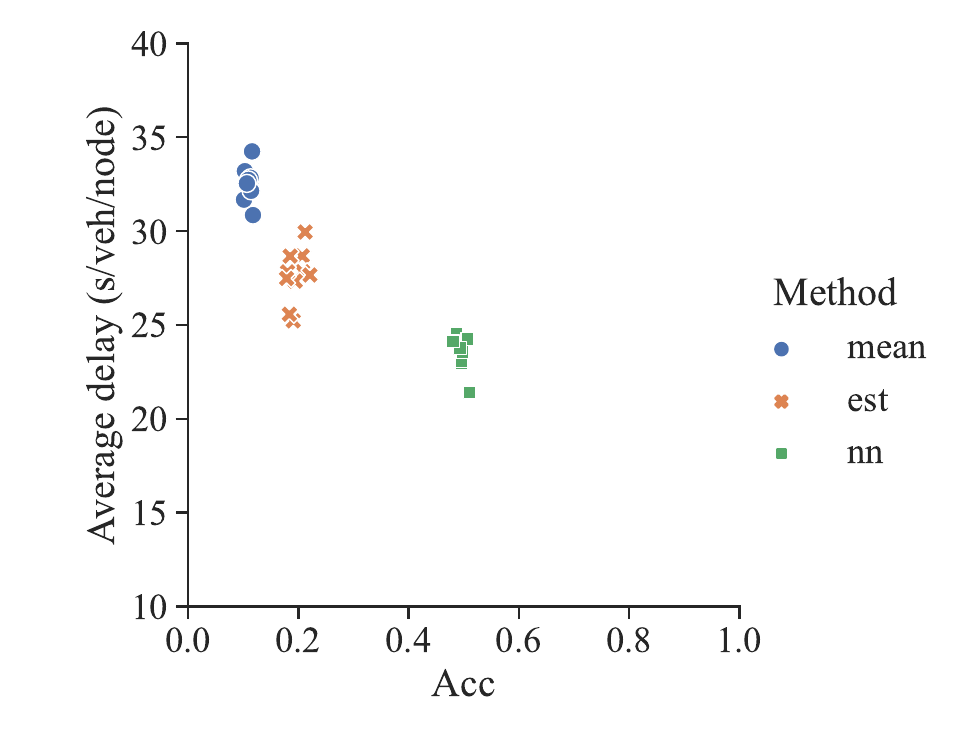}}
		\label{F:lescbp_delay1.5}}
	\caption{Delay and accuracy for LESCBP control.
	}
	\label{F:lescbp_delay}
\end{figure}

Similar to \autoref{F:pwbp_delay}, we show the result of LESCBP when demand is set to 100\% and 150\% levels of field demand in \autoref{F:lescbp_delay} \subref{F:lescbp_delay1} and \subref{F:lescbp_delay1.5}, respectively. We can find a similar performance of LESCBP compared with PWBP and an apparent trend that the delay of LESCBP decreases with an increase in accuracy. Therefore, we conclude that more knowledge of I-SFR can help LESCBP reduce vehicle delay. 

The results of scenarios with only cars (the same simulation network without non-motorized vehicles and buses) appear in \ref{Ap:delay}. Different methods still produce significantly different accuracies, and the finding that higher prediction ability helps reduce delay is still valid in this purer network. These indicate that our theories are universal for different traffic environments.

\begin{figure}[h!]
	\centering
	\resizebox{0.85\textwidth}{!}{
		\includegraphics{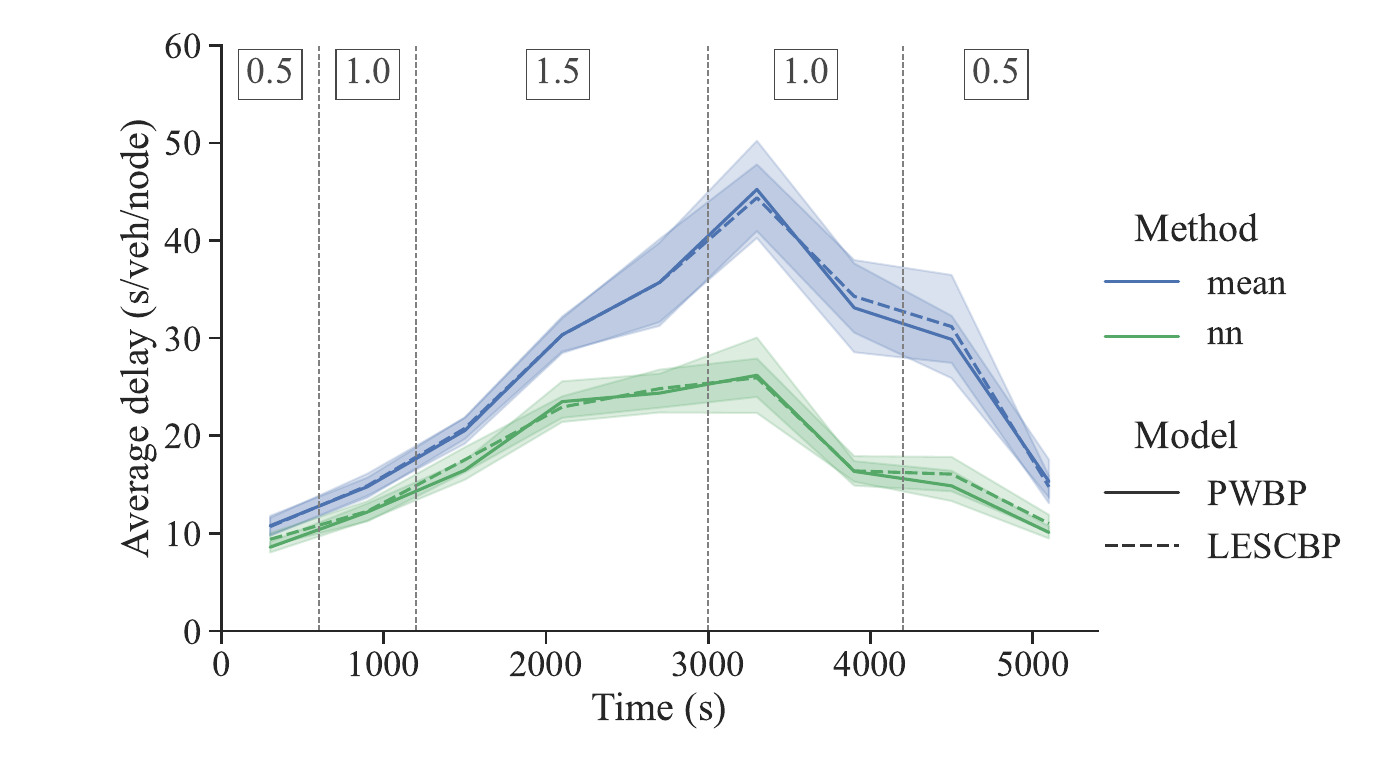}}
	\caption{Average delay under different control models and I-SFR prediction methods.} 
	\label{F:delay_vary}
\end{figure}

Finally, we show comparisons between ``mean'' and ``nn'' and between PWBP and LESCBP with varying demand in \autoref{F:delay_vary}. The simulation lasts for 1.5 h. The initial demand is 50\% of the field demand, which lasts for 10 min. It then increases to 100\% and 150\% field demand gradually. After 30 min of 150\% field demand, it gradually recovers to 100\% and 50\% field demand. Each simulation is run ten times with different random seeds, and \autoref{F:delay_vary} shows the mean and standard deviation of the average vehicle delay. The solid line represents PWBP, while the dashed line represents LESCBP. Different colors represent different prediction methods. We first focus on the comparison between mean-PWBP and mean-LESCBP. Similar to the findings in \cite{wang2022learning}, we can see that 1) when the demand is low, mean-PWBP performs slightly better than mean-LESCBP; 2) when the demand becomes high, mean-LESCBP starts to outperform mean-PWBP. These findings, which are in good agreement with the results in \cite{wang2022learning}, imply that we reproduced mean-LESCBP of \cite{wang2022learning}. We then focus on two prediction methods. Clearly, in terms of average delay, ``nn'' outperforms ``mean'' a lot during the whole period for both PWBP and LESCBP. This validates that increasing the I-SFR's prediction accuracy helps reduce the network delay under varying demands. 

\section{Conclusion and outlook}
\label{S:Conc}
The size of a network's stability region reflects the urban network's ability to handle traffic demands. In this paper, we abandon the assumption of completely known saturation flow rate (SFR) in existing real-time traffic control studies and analyze how the knowledge of imminent SFR (I-SFR) affects network supply.
We analytically prove that improving the I-SFR's prediction accuracy can enlarge the size of stability region. We demonstrate that, as long as the demand rates are interior to the corresponding stability region, the BackPressure (BP) with predicted I-SFR can guarantee network stability. Results from simulations validate the proposed theory on one hand and show the improvements in network delay with more knowledge of I-SFR on the other hand.

BP-based controls serve as the ideal tool to detect the changes in network's stability region since they can maximize the network throughput. However, this does not mean the knowledge of I-SFR only benefits BP-based control algorithms. Future research can investigate the influence of I-SFR on other real-time control policies. The key point is to realize that the global supply of traffic network is not fixed and can be influenced by knowledge of real-time local supply. The three prediction methods in the simulation mainly aim to generate different levels of I-SFR. We hence did not focus on improving the prediction accuracy to the maximum degree. Although we show the advantages of the neural network prediction method over the mean I-SFR method and the heuristic method through experiments, we must point out that any prediction method has its price. A prediction method with a higher accuracy may require a more advanced hardware facility, which usually comes with greater economic investment. In the near future, when the neural network method (or other similar methods) is not feasible despite its high accuracy, it is the mean I-SFR or heuristic method that works in practice. However, we need to keep in mind that efforts to improve prediction accuracy to the extent that technology allows is worthwhile.

\section*{Acknowledgments}
\label{Ack}
This work was supported by the National Natural Science Foundation of China under Grants 52302406 and 72201065, and by the Natural Science Foundation of Fujian Province under Grants 2023J05022 and 2023J05099.

\newpage
\appendix
\section{Table of notations}
\label{Ap:notations}
\vspace{-12pt}
\begin{table}[h!]
	\centering
	\setlength{\extrarowheight}{-1.5pt}
	\label{T:Notation}
	\begin{tabular} {c p{0.8\linewidth}} 
		\hline
		\\[-0.8em]
		\textbf{Parameter} & \textbf{Description} \\
		\\[-0.9em]
		\hline
		\\[-0.8em]	
		$\subset$, $\subseteq$ & left set is a strict subset of right set, left set is a subset of or equal to right set\\
		$\times$ or $\bigtimes$ & Cartesian product \\
		$\vdash$ & a point (left) dominates another point (right) \\
		$\mathbb{1}\{\cdot\}$ & indicator function which is equal to 1 if true and 0 otherwise\\
		$a$, $\widetilde{a}$, $\bm{a}$ & exogenous arrival rate, upper bound of $a$, exogenous arrival rate vector \\
		$c$, $\bm{c}$ & capacity, capacity vector \\ 
		$\mathcal{D}$, $\mathcal{D}^-$ & stability region, stability region excluding the general upper frontier \\
		$d(v)$ & distance from most-upstream point of $v$'s located link to $v$'s current location \\
		$\mathcal{E}$ & error, equal to the difference between I-SFR's real value and prediction value \\
		$E$, $|E|$, $e$ & set of joint I-SFR values: $E = \bigtimes_{m \in \MM} S_m$, size of $E$, a joint value: $e \in \{1,2,\cdots,|E|\}$ \\
		$F(\cdot)$, $\psi$  & switching curve in LESCBP, switch function in LESCBP \\
		$\bar{f}(\cdot)/f(\cdot)$ & precise/general upper frontier of the region ``$\cdot$''\\ 
		$g$, 
		$\bm{g}$, $\mathcal{G}$ & (effective) green ratio, 
		green ratio vector, $\bm{g}$'s complete set  \\
		$\bm{h}$ & (column) constraint vector of signalized intersection \\
		$i$/$j$ & $m$'s upstream/downstream movement \\
		$K$, $\widetilde{K}$ & positive constants \\
		$\bm{K}$ & constraint matrix for all conflicting movements and 0-boundaries \\
		$\LX$ & set of all directed links\\
		$L(\cdot)$, $\Delta(\cdot)$ & network-wide Lyapunov function, Lyapunov drift function \\
		$l(m)$ & length of the link that $m$ originates from\\
		$\MM$, $\MM_n$, $|\MM|$ & set of all movements, set of $n$'s movements, number of all movements\\
		$m$ (or $\mu$) & a movement \\
		$n$, $\NX$, $|\NX|$ & a node, set of all network nodes, number of all network nodes \\
		$O(m)$ & set of $m$'s all downstream movements \\
		$p^e$, $\hat{p}$ & probability of the $e$th joint value, probability measure in guess\\
		$q_m$ & actual departure of $m$ \\
		$r_m$, $\bm{R}$ & average turning ratio of movement $m$, average turning ratio matrix \\ 
		$S_m$ & set of all I-SFR values in $m$\\ 
		$s$, $\overline{s}$, $\hat{s}$, $\widetilde{s}$, $\bm{s}$ & I-SFR, M-SFR, estimated I-SFR, upper-bound of I-SFR, I-SFR vector \\
		$t$, $T$ & time of decision, upper bound of time \\
		$U(m)$  & set of $m$'s all upstream movements \\
		$v$, $V_m$ & a vehicle, all vehicles queueing in $m$ \\
		$w$ & weight in BP control\\
		$x_m$, $\bm{x}_n$, $\bm{X}$ & $m$'s queue length, queue length vector of $n$, queue length vector of network  \\
		$y,Y$ & an I-SFR value type number, total number of I-SFR value types in a movement\\ 
		$\epsilon$, $\bm{\epsilon}$ & reserve demand calculated by SFR-only algorithm, reserve demand vector  \\
		$\theta$, $\eta$ & ability of I-SFR prediction, accuracy of I-SFR prediction \\
		$\lambda_m$, $\bm{\lambda}$ & $m$'s demand arrival rate consisting of exogenous demand and upstream demands, demand arrival rate vector \\
		$\rho_m$ & different I-SFR value's probability of $m$\\
		$\bm{\Phi}_n$, $\bm{\Phi}$  & set of allowable phases at $n$, set of allowable network phases: $\bm{\Phi} \equiv \bigtimes_{n \in \NX} \bm{\Phi}_n$ \\
		$\Phi_n$, $\Phi$ & allowable phases at $n$: $\Phi_n \in \bm{\Phi}_n$, allowable network phases: $\Phi \equiv \bigtimes_{n \in \NX} \Phi_n \subset \bm{\Phi}$ \\
		$\phi_m$, $\bm{\phi}$ & $m$'s passing status (1 if passing is allowed, and 0 otherwise), the vector of passing status of all movements\\
		\\[-0.9em]				
		\hline
	\end{tabular}
	\label{t:notation}
\end{table}

\section{Proofs}
\subsection{Proof of Proposition 2}
\label{Ap:P2}
\begin{proof}
	$\forall (\bm{a},\bm{R}): \bm{0} \leq (\bm{I} - \bm{R})^{-1}\cdot\bm{a} \leq \bm{\overline{s}}\cdot \bm{g}$, where $g \in \GX$, $\exists$ diagonal matrix $\bm{\gamma}$ (all diagonal elements are between 0 and 1) to let $(\bm{I} - \bm{R})^{-1}\cdot\bm{a} = \bm{\gamma}\cdot\bm{\overline{s}}\cdot \bm{g} = \bm{\overline{s}} \cdot (\bm{\gamma} \cdot \bm{g})$. Because $(\bm{\gamma} \cdot \bm{g}) \in\GX$, we can let $\bm{g}' \equiv \bm{\gamma} \cdot \bm{g}$, and hence $(\bm{a},\bm{R}) \in \{(\bm{a}',\bm{R}'):(\bm{I} - \bm{R}')^{-1}\cdot\bm{a}' = \bm{\overline{s}} \cdot \bm{g}',~~ \bm{g}' \in \GX\}$. Besides, it is clear that $\forall (\bm{a'},\bm{R}'): (\bm{I} - \bm{R}')^{-1}\cdot\bm{a}' = \bm{\overline{s}}\cdot \bm{g}'$, where $\bm{g}'\in \GX$, we have $(\bm{a}',\bm{R}') \in \{ (\bm{a},\bm{R}): \bm{0}\leq(\bm{I} - \bm{R})^{-1}\cdot\bm{a}\leq\bm{\overline{s}}\cdot\bm{g},~~ \bm{g} \in \GX\}$. Hence two spaces are the same.
\end{proof}

\subsection{Proof of Proposition 3}
\label{Ap:P3}
\begin{proof}
	For a movement $m$, suppose it has $Y=|\SX_m|$ kinds of I-SFR values. We denote the $y$th I-SFR as $\SX_m^y$, $y = 1,2,\cdots, Y$ and its probability as $\rho_m^y$. Clearly, $\sum_{y=1}^Y \rho^y_m = 1$.
	Furthermore, we denote the guessed probability for the $y$th I-SFR as $\hat{p}_m^y$. Note that the guessed I-SFR may be outside of set ${\SX_m}$, there is $ 0<\sum_{y=1}^Y \hat{p}_m^y \leq 1$. If we allow a guess error and regard the guessed I-SFR within the error as a correct guess, hence
	$\hat{p}_m^y = \int_{\hat{s}_m}\mathbb{1}\{\SX_m^y-b/2 \leq \hat{s}_m < \SX_m^y+b/2\} \dd \hat{p}(\hat{s}_m)$, where $\hat{s}_m$ is the guessed I-SFR, $\mathbb{1}\{\cdot\}$ is indicator function equal to 1 if true and 0 otherwise, and $b/2$ is the allowed error in guess. $b$ should not be larger than the interval between two adjacent I-SFRs. If $b=0$, the estimated accuracy does not allow any error. 
	Given $\theta$, $m$'s I-SFR prediction accuracy, denoted by $\eta$, can be calculated as: 
	\begin{equation}
		\eta|_\theta 
		= \theta + (1-\theta) \sum_{y=1}^Y \rho_m^y \cdot \hat{p}_m^y. 
	\end{equation}
	Because $\theta \in [0,1]$ and
	$\sum_{y=1}^Y \rho_m^y \cdot \hat{p}_m^y< \sum_{y=1}^Y \rho_m^y \cdot \sum_{y=1}^Y \hat{p}_m^y \leq 1$ when $Y\geq 2$, we have that $\eta$ increases monotonically with $\theta$, and vice versa.
\end{proof}

\subsection{Proof of Lemma 2}
\label{Ap:L2}
\begin{proof}
	The capacity under the SFR-only strategy is $\bm{c}|_\theta = \EE \{ \bm{s}(t)\cdot\bm{\phi}^{\mathrm{SFR-only}}(t) \} = \theta \cdot \sum_{e=1}^{|E|}p^e \cdot \bm{s}^e \cdot \bm{g}^e  + (1-\theta) \cdot \overline{\bm{s}} \cdot 
	\bm{g}$. As $\EE \{\bm{R}(t)\} = \bm{R}$, $\EE \{ \bm{a}(t) \} = \bm{a}$, from the constraint in (\red{18}), we have 
	\begin{equation}
		\EE \{ \bm{I} - \bm{R}(t) \} \cdot \ \EE \{ \bm{s}(t)\cdot\bm{\phi}^{\mathrm{SFR-only}}(t) \} -\EE \{ \bm{a}(t) \} \geq \bm{\epsilon}.
	\end{equation}
	Since $\bm{R}(t)$ is not observed by SFR-only, they are independent and hence we have 
	\begin{equation}
		\EE \{(\bm{I}-\bm{R}(t)
		) \cdot \bm{s}(t) \cdot \bm{\phi}^{\mathrm{SFR-only}}(t) - \bm{a}(t) \} \geq \bm{\epsilon}. 
	\end{equation}
	When the demand rate lies in $\mathcal{D}_{\theta}^-$, we have $\epsilon_{\max} > 0$. Hence by setting $\epsilon = \epsilon_{\max}$, we have 
	\begin{equation}
		\EE \{(\bm{I}-\bm{R}(t)) \cdot \bm{s}(t) \cdot \bm{\phi}^{\mathrm{SFR-only}}(t) - \bm{a}(t) \} \geq \bm{\epsilon_{\max}} > \bm{0}. 
	\end{equation}
\end{proof}

\subsection{Proof of Lemma 3}
\label{Ap:L3}
\begin{proof}
	For each $t \ge 0$ we have 
	\begin{multline} 
		\EE \big\{\hat{s}_m(t)\phi_m(t) - a_m(t) - r_m(t)\sum_{i \in U(m)}  \hat{s}_i(t)\phi_i(t)  \big| \bm{X}(t)\big\}
		\\ = \EE \{s_m(t)\phi_m(t)\big| \bm{X}(t)\} + \EE \{\mE_m(t)\phi_m(t)\big| \bm{X}(t)\} - \EE \{a_m(t) \}
		~~~~~~~~~~~~~~~~~~~~~~~~~~~~\\ ~~~~~~~~~~~~~~~~~~~~~~-\EE\{r_m(t) \sum_{i \in U(m)}  s_i(t) \phi_i(t) \big| \bm{X}(t)\}
		- \sum_{i \in U(m)}  \EE\{r_m(t)\mE_i(t)\phi_i(t)\big| \bm{X}(t)\}
		\label{E:hat(t)}
	\end{multline}
	Note that $\phi_m(t) = 1~\mathrm{or}~0$, we have 
	\begin{equation}
		\EE \{\mE_m(t)\phi_m(t)\big| \bm{X}(t)\} = \EE \{\mE_m(t)\big| \bm{X}(t), \phi_m(t)=1 \} \PP\{ \phi_m(t)=1 \} + 0,
	\end{equation}
	and
	\begin{equation}
		\EE \{r_m(t)\mE_i(t)\phi_i(t)\big| \bm{X}(t)\} = \EE \{r_m(t)\mE_i(t)\big| \bm{X}(t), \phi_i(t)=1 \} \PP\{ \phi_i(t)=1 \} + 0
	\end{equation}
	for $i \in U(m)$.
	$r_m(t)$ is the proportion of all vehicles from $m$'s upstream movements $U(m)$ that will directly join movement $m$ at time $t$, hence it only influences the turning ratio of $m$ at time $t+1$, and can hardly influence the prediction error of $m$'s I-SFR at time $t$. $r_m(t)$ is also not influenced by $\mE_m(t)$ logically. Therefore, $\mE_i(t)$ and $r_m(t)$ are independent. 
	Given $\phi_m = 1$ for any $m$, $\mE_m(t)$ is also independent of $x_m(t)$ (which is analyzed in detail in \autoref{Ap:independence}), we have
	
	\begin{equation}
		\EE \{\mE_m(t)\big| \bm{X}(t), \phi_m(t)=1 \} = \EE \{\mE_m(t)\big|\phi_m(t)=1 \} = 0,
	\end{equation}
	and
	\begin{equation}
		\EE \{r_m(t)\mE_i(t)\big| \bm{X}(t), \phi_i(t)=1 \} = \EE \{r_m(t)\big| \bm{X}(t), \phi_i(t)=1 \} \EE \{\mE_i(t)\big| \phi_i(t)=1 \} = 0.
	\end{equation}
	Hence, \eqref{E:hat(t)} is equal to
	\begin{equation}
		\EE \big\{s_m(t)\phi_m(t) - a_m(t) - r_m(t)\sum_{i \in U(m)}  s_i(t)\phi_i(t) \big| \bm{X}(t)\big\}.
	\end{equation}
\end{proof}

\subsection{Proof of Lemma 4}
\label{Ap:L4}
\begin{proof}
	We first integrate both sides of (\red{27}) over the interval $[0,T]$ and take expectation of both sides of the inequality to obtain
	\begin{equation}
		\EE \sum_{t=0}^{T-1}\EE \{ L \big( \boldsymbol{X}(t+1) \big) -  L\big( \boldsymbol{X}(t) \big) | \boldsymbol{X}(t) \}  \le KT - \epsilon \Big(\sum_{t=0}^{T-1} \sum_{m \in \MM} \EE \{x_m(t)\} \Big).	\label{temp1}
	\end{equation}
	Using the law of iterated expectations and telescoping sums yields
	\begin{equation}
		\EE \{ L \big( \boldsymbol{X}(T-1) \big)\} -  \EE \{L\big( \boldsymbol{X}(0) \big) \} \le KT - \epsilon \Big(\sum_{t=0}^{T-1} \sum_{m \in \MM} \EE \{x_m(t)\} \Big).
	\end{equation}
	Rearranging terms, dividing by $\epsilon T$, and using the fact that $L\big( \boldsymbol{X}(T-1) \big) > 0$, we have
	\begin{equation}
		\frac{1}{T} \sum_{t=0}^{T-1} \sum_{m \in \MM} \EE \{x_m(t)\} < \frac{K}{\epsilon} + \frac{\EE \{L\big( \boldsymbol{X}(0) \big) \}}{\epsilon T}.
	\end{equation}
	Since $\EE \{L \big( \boldsymbol{X}(0) \big) \} < \infty$, taking a lim sup yields:
	\begin{equation}
		\underset{T \rightarrow \infty}{\lim \sup} ~ \frac{1}{T} \sum_{t=0}^{T-1} \sum_{m \in \MM} \EE \{x_m(t)\} < \frac{K}{\epsilon}.
	\end{equation}
\end{proof}

\subsection{Proof of Theorem 2}
\label{Ap:T2}
\begin{proof}
	Substituting (\red{3}) 
	into (\red{25}) 
	, and noting that $ \max[x_m(t) - s_m(t)\phi_m(t), 0] \leq x_m(t) $, $\max[x_m(t) - s_m(t)\phi_m(t), 0]^2 \leq [x_m(t) - s_m(t)\phi_m(t)]^2 $, and 
	$\min[x_m(t), s_m(t)\phi_m(t)] \leq s_m(t)\phi_m(t) $, we have: 
	\begin{multline}
		L \big( \boldsymbol{X}(t+1) \big) \leq \frac{1}{2}\sum_{m \in \MM} 
		\big\{ \big[x_m(t) - s_m(t)\phi_m(t)\big]^2 \\ + \big[a_m(t) + r_m(t)\sum_{i \in U(m)}  s_i(t)\phi_i(t)\big]^2 + 2x_m(t)\big[a_m(t) + r_m(t)\sum_{i \in U(m)}  s_i(t)\phi_i(t)\big] \big\}.
		\label{E:L(X(t+1))}
	\end{multline}
	Substituting \eqref{E:L(X(t+1))} into (\red{26}) 
	yields:
	\begin{multline}
		\Delta \big( \boldsymbol{X}(t) \big) \leq \EE \Big\{ \sum_{m \in \MM} \big\{ -x_m(t)s_m(t)\phi_m(t) +\frac{1}{2}[s_m(t)\phi_m(t)]^2
		+ \frac{1}{2}[a_m(t) \\+ r_m(t)\sum_{i \in U(m)}  s_i(t)\phi_i(t)]^2 + x_m(t)\big[a_m(t) + r_m(t)\sum_{i \in U(m)}  s_i(t)\phi_i(t)\big] \big\} \big| \bm{X}(t)\Big\}.
		\label{E:drift1}
	\end{multline}
	Because $[s_m(t)\phi_m(t)]^2 + \big[a_m(t) + r_m(t)\sum_{i \in U(m)}  s_i(t)\phi_i(t)\big]^2 \leq \widetilde{s}^2 + \underset{m}{\max}[\widetilde{a} + \sum_{i \in U(m)}\widetilde{s}]^2 $ for $m \in \MM$, where $\widetilde{s}$
	is the maximum value of the I-SFR of all possible movements, and $\widetilde{a}$ 
	is the maximum value of all possible exogenous arrival rates. Let $\widetilde{K} = \widetilde{s}^2 + \underset{m}{\max}[\widetilde{a} + \sum_{i \in U(m)}\widetilde{s}]^2$, the second term plus the third term on the right-hand side of \eqref{E:drift1} is then bounded by $\frac{1}{2}|\MM|\widetilde{K}$.
	Rearranging terms of \eqref{E:drift1} and utilizing the properties of conditional expectation, we have
	\begin{multline}
		\Delta \big( \boldsymbol{X}(t) \big) \leq  
		\frac{1}{2}|\MM|\widetilde{K} - \sum_{m \in \MM} x_m(t) \EE \big\{s_m(t)\phi_m(t) - a_m(t) - r_m(t)\sum_{i \in U(m)}  s_i(t)\phi_i(t)  \big| \bm{X}(t)\big\},
		\label{E:drift2}
	\end{multline}
	where $x_m(t)$ is moved out of the expectation since $\bm{X}(t)$ is given.
	
	Define a control policy $\Phi^{\max}$ for each $t \ge 0$ as 
	\begin{equation}
		\Phi^{\max} = \underset{\Phi \in \bm{\Phi}}{\arg\max} \Big( \sum_{m \in \MM} x_m(t) \EE \big\{\hat{s}_m(t)\phi_m(t) - a_m(t) - r_m(t)\sum_{i \in U(m)}  \hat{s}_i(t)\phi_i(t)  \big| \bm{X}(t)\big\} \Big).
		\label{E:pmax}
	\end{equation}
	
	By definition and combining with Lemma 3, we have for each $t \ge 0$ that 
	\begin{multline}
		\sum_{m \in \MM} x_m(t) \EE \big\{\hat{s}_m(t)\phi_m^{\max}(t) - a_m(t) - r_m(t)\sum_{i \in U(m)}  \hat{s}_i(t)\phi_i^{\max}(t) \big| \bm{X}(t)\big\} \\ \geq \sum_{m \in \MM} x_m(t) \EE \big\{s_m(t)\phi_m^*(t) - a_m(t) - r_m(t)\sum_{i \in U(m)}  s_i(t)\phi_i^*(t) \big| \bm{X}(t)\big\}.
		\label{E:themPf1}
	\end{multline}
	where $\phi_m^{\max}(t)$ represent the passing status of $m$ at time $t$ when $\Phi^{\max}$ is implemented, and $\phi_m^*(t)$ represents the passing status of $m$ at time $t$ when any other particular policy $\Phi^*$ (including the SFR-only policy) is implemented. Setting $\Phi^* = \Phi^{\mathrm{SFR-only}}$ and plugging \eqref{E:themPf1} into \eqref{E:drift2}, we have
	\begin{equation}
		\Delta \big( \boldsymbol{X}(t) \big) \le \frac{1}{2}|\MM|\tilde{K} - \sum_{m \in \MM} x_m(t) \EE\{s_m(t)\phi^{\mathrm{SFR-only}}_m(t) - a_m(t) - r_m(t)\sum_{i \in U(m)}s_i(t)\phi^{\mathrm{SFR-only}}_i(t)\}, 
		\label{drift3}
	\end{equation}
	where the left-hand-side represents the drift under $\Phi^{\max}$, and $\boldsymbol{X}(t)$ is dropped from the expectation since the SFR-only algorithm is independent of $\boldsymbol{X}(t)$.
	
	Since we assume the demand rate lies in $\mathcal{D}_{\theta}^-$, we have from \red{Lemma 2} that there exists an $\epsilon > 0$ such that
	\begin{equation}
		\sum_{m \in \MM} x_m(t) \EE\{s_m(t)\phi^{\mathrm{SFR-only}}_m(t) - a_m(t) - r_m(t)\sum_{i \in U(m)}s_i(t)\phi^{\mathrm{SFR-only}}_i(t)\} \geq \epsilon \sum_{m \in \MM} x_m(t)
		\label{E:themPf0}
	\end{equation}	
	Substituting \eqref{E:themPf0} into \eqref{drift3}, setting $K \equiv \frac{1}{2}|\MM|\tilde{K}$, we have by appeal to \red{Lemma 4} that $\Phi^{\max}$ is also network stabilizing. 
	
	It remains to show that solving \eqref{E:pmax} ($\Phi^{\max}$) is equivalent to solving (\red{23}) (the BP policy with predicted I-SFR). \eqref{E:pmax} can be rewritten as
	\begin{equation}
		\Phi^{\max} = \underset{\Phi \in \bm{\Phi}}{\arg\max} \Big(\sum_{m \in \MM} x_m(t) \hat{s}_m(t)\phi_m(t) - \sum_{m \in \MM} x_m(t)r_m(t)\sum_{i \in U(m)}\hat{s}_i(t)\phi_i(t)\Big).
		\label{E:opt1}
	\end{equation}
	The term corresponding to exogenous arrivals in \eqref{E:pmax} was dropped from the optimization problem since it constitutes an additive constant to the problem.  We have also applied the principle of \textit{opportunistically maximizing an expectation} to drop the expectations in \eqref{E:pmax} from the problem.
	Re-arranging the orders of summation, \eqref{E:opt1} becomes
	\begin{equation}
		\Phi^{\max} = \underset{\Phi \in \bm{\Phi}}{\arg \max}\sum_{m \in \MM} \hat{s}_m(t)\phi_m(t)\big[x_m(t) - \sum_{j \in O(m)}x_j(t)r_j(t)\big]
		\label{E:opt3}
	\end{equation}
	Using the definition of $w^{\mathrm{BP}}_m(t)$, \eqref{E:opt3} becomes
	\begin{equation}
		\Phi^{\max} = \underset{\Phi \in \bm{\Phi}}{\arg \max} \sum_{m \in \MM} w_m^{\mathrm{BP}}(t)\hat{s}_m(t)\phi_m(t)
		\label{E:opt4}
	\end{equation}
	Since intersection movements do not interact across nodes instantaneously and $\Phi_{n_1} \cap \Phi_{n_2} = \emptyset$ for any $n_1,n_2 \in \NX$ such that $n_1 \ne n_2$, the optimization problem naturally decomposes by intersection.  That is, maximizing \eqref{E:opt4} is equivalent to solving the $|\NX|$ problems
	\begin{equation}
		\Phi^{\mathrm{BP}}_n(t) \in \underset{\Phi_n \in \bm{\Phi}_n}{\arg \max} \sum_{m \in \MM_n} w_m^{\mathrm{BP}}(t) \hat{s}_m(t)\phi_m(t), \quad \forall n \in \NX. \label{E:BPn}
	\end{equation}
	Hence, solving \eqref{E:pmax} turns out to be the same as solving (\red{23}).
\end{proof}

\section{Demand patterns}
\label{Ap:demand}
\begin{figure}[!ht]
	\centering
	\subfloat[][Motorized vehicles]{\resizebox{0.58\textwidth}{!}{
			\includegraphics[width=0.4\textwidth]{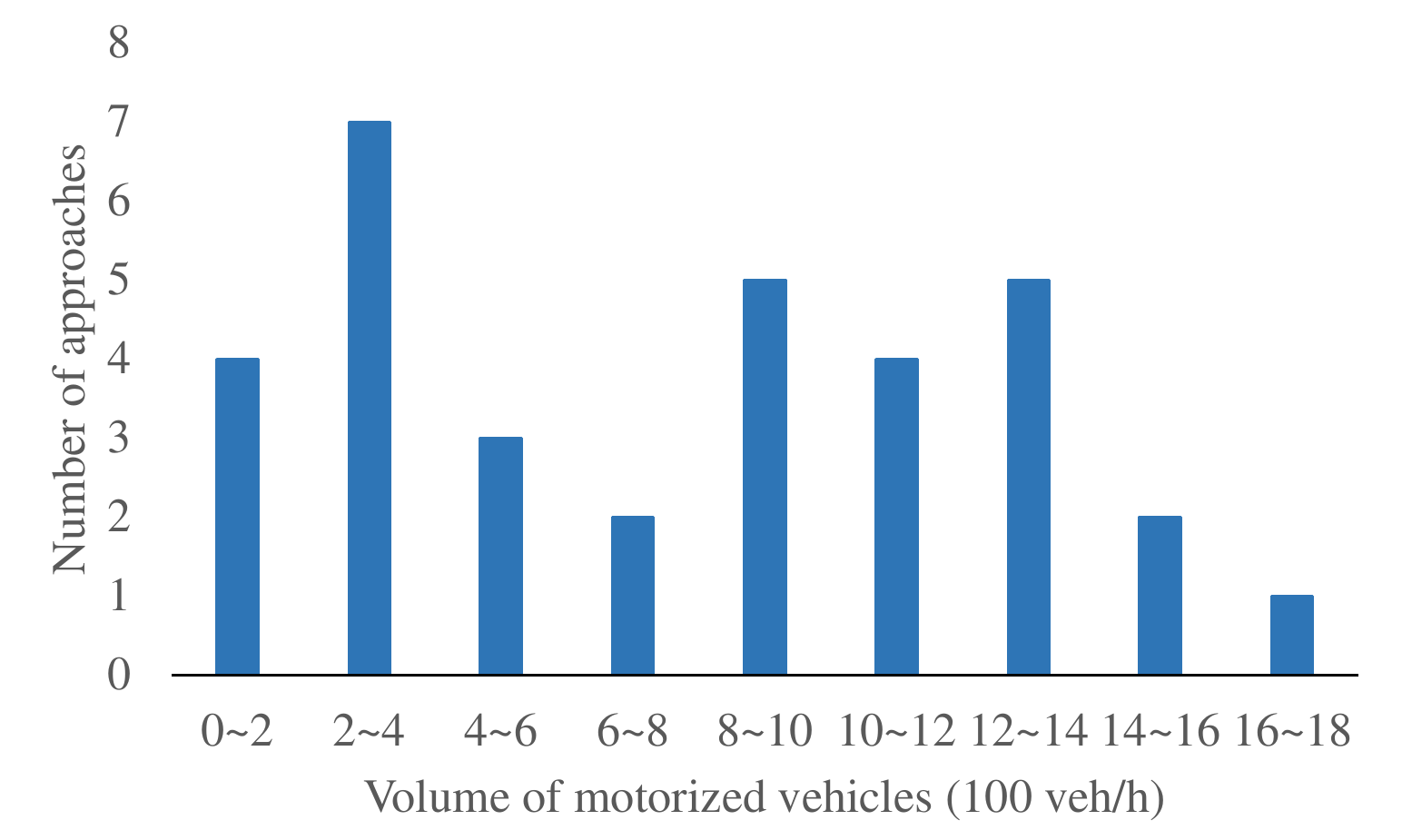}}
		\label{F:mv}} 
	
	\subfloat[][Non-motorized vehicles]{\resizebox{0.58\textwidth}{!}{
			\includegraphics[width=0.4\textwidth]{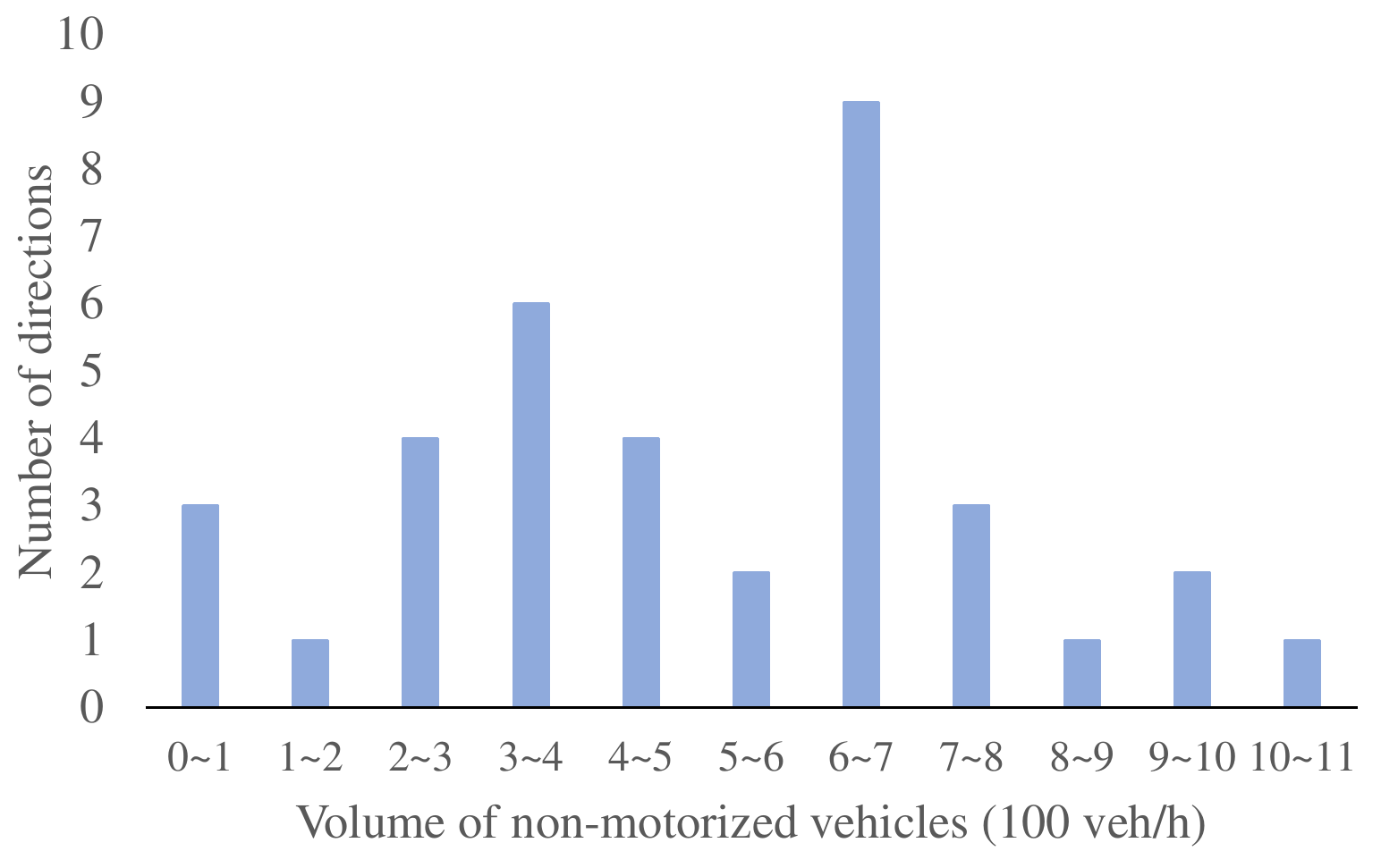}}
		\label{F:nmv}}
	
	\subfloat[][Bus distribution]{\resizebox{0.58\textwidth}{!}{
			\includegraphics[width=0.4\textwidth]{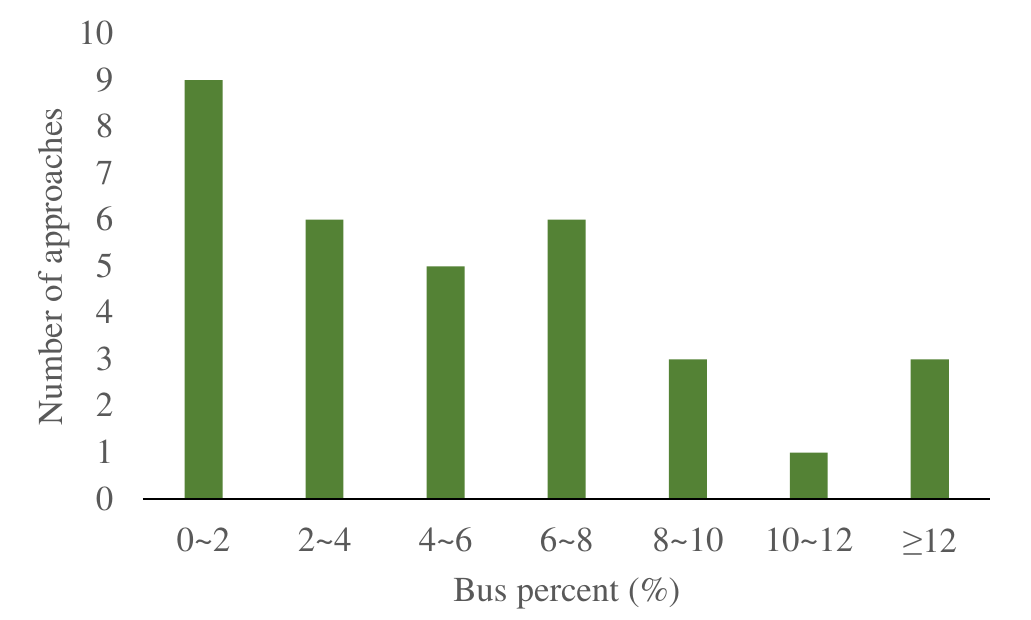}}
		\label{F:bp}}
	\caption{Demand patterns of peak hour along Qishan avenue.}
	\label{F:demand}
\end{figure}

The vehicular demands patterns collected during the evening peak hour are shown \autoref{F:demand}.

\section{Choice of threshold for the reserve demand} 
\label{Ap:choice}
\begin{figure}[h!]
	\centering
	\resizebox{0.52\textwidth}{!}{
		\includegraphics{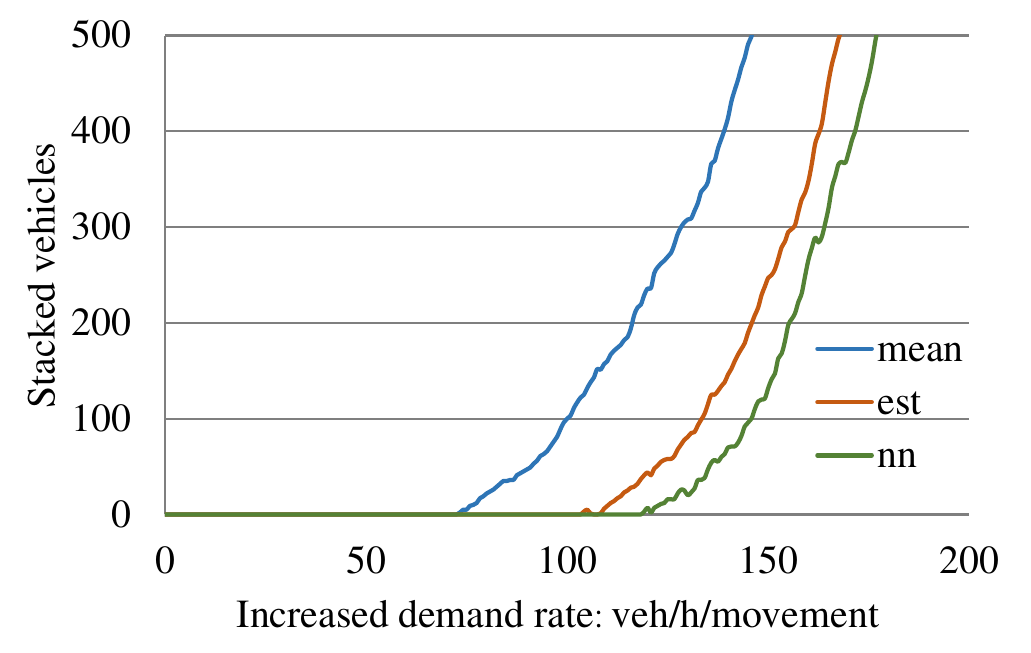}}
	\caption{Stacked vehicle of the network with the increase of demand rate given a random seed.} 
	\label{F:s_v}
\end{figure}

In this appendix, we take BP as an example to show why 100-veh is acceptable as the threshold for determining reserve demand. \autoref{F:s_v} shows the number of stacked vehicles with the increased demand rate given a fixed random seed.

\begin{figure}[h!]
	\centering
	\resizebox{0.52\textwidth}{!}{
		\includegraphics{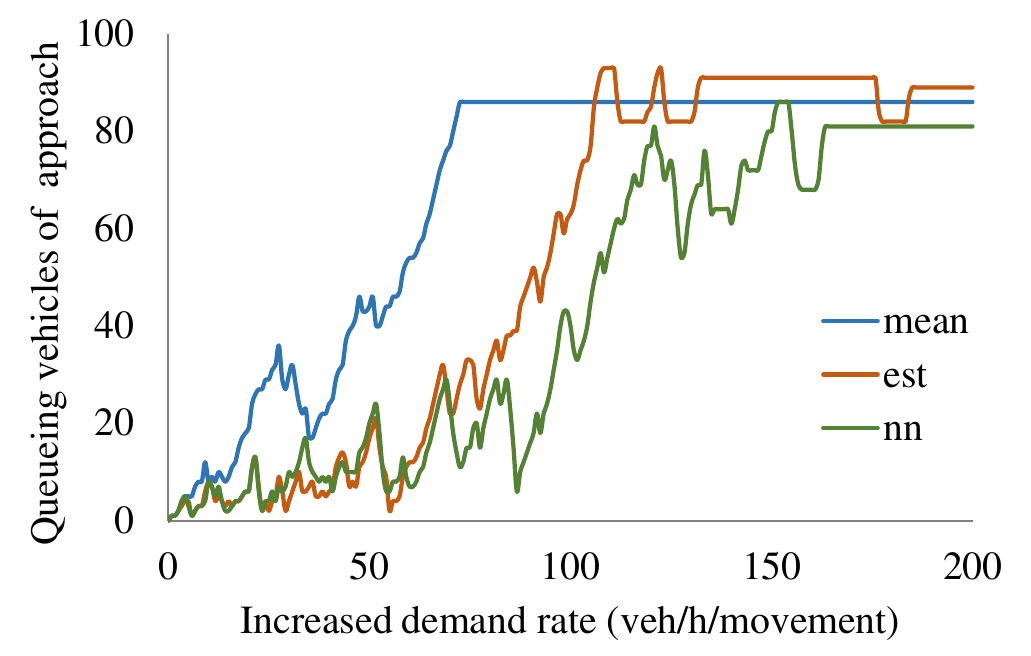}}
	\caption{Queueing vehicles at approach { \textcircled{\smaller 1}}-west with increase of demand rate given a random seed.} 
	\label{F:q_v}
\end{figure}

The curves of staked vehicles include two parts. From 0 to about 100 increased demand rate (about 80 for ``mean'' and 120 for ``nn''), the number of stacked vehicles is about 0, which means the demand is within the stability region. When the increased demand rate is larger than 100, the number of stacked vehicles keeps growing, which means demand is going beyond the stability region. However, it is hard to judge at which moment or stacked-vehicle number the demand exactly reaches the frontier of stability region. Despite this, we can find that once the number of stacked vehicles increases, the relative positions of the three curves are stable. Setting the threshold to 100, 200, or 300 vehs makes little difference to the gap between different methods' reserve demand. Therefore, we choose 100-veh as the threshold as 1) it is big enough to represent that the demand has reached the general upper frontier of stability region, and 2) it is not too big and can save simulation time.

In \mbox{\autoref{F:q_v}}, we further display the queueing vehicles (sum of queueing lengths for all movements in an approach) in a critical approach (easy to spill back) given the same random seed. Note that a curve's ``platform stage'' means the whole approach is blocked, and no vehicles can enter. Because vehicles' sizes and space headways differ, the vehicle numbers at the ``platform stage'' could also vary. Once the ``platform stage'' is reached, no vehicles can pass the intersection (including the right-turns since they are blocked by the through and left-turn vehicles). In general, ``est'' perform better than ``mean'' because it arrives at the ``platform stage'' later. And ``nn'' performs even better than ``est''. This is consistent with the results in \red{Figure 12} and \autoref{F:s_v}.

\section{Delay in a simulation only with cars}
\label{Ap:delay}
\begin{figure}[H]
	\centering
	\subfloat[][BP]{\resizebox{0.32\textwidth}{!}{
			\includegraphics[width=0.5\textwidth]{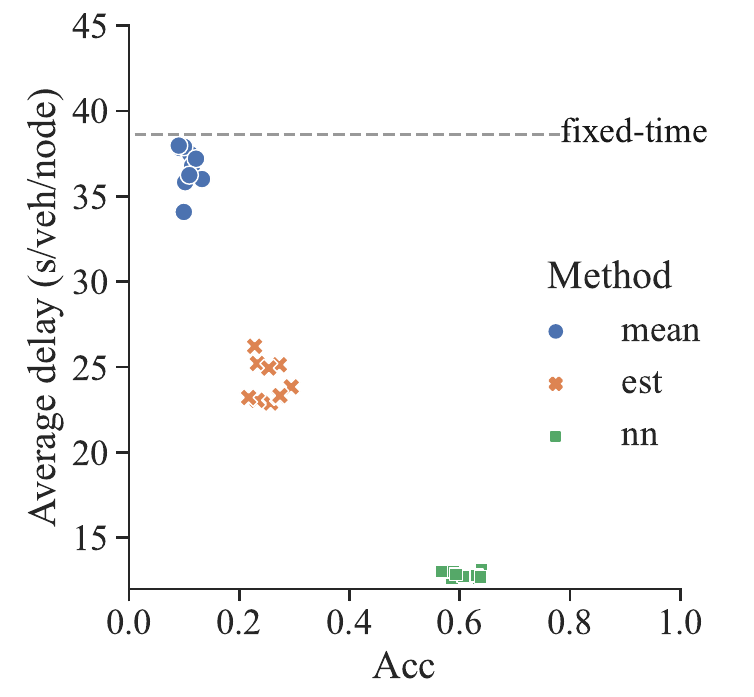}}
		\label{F:bp0_delay}} 
	~~
	\subfloat[][PWBP]{\resizebox{0.32\textwidth}{!}{
			\includegraphics[width=0.4\textwidth]{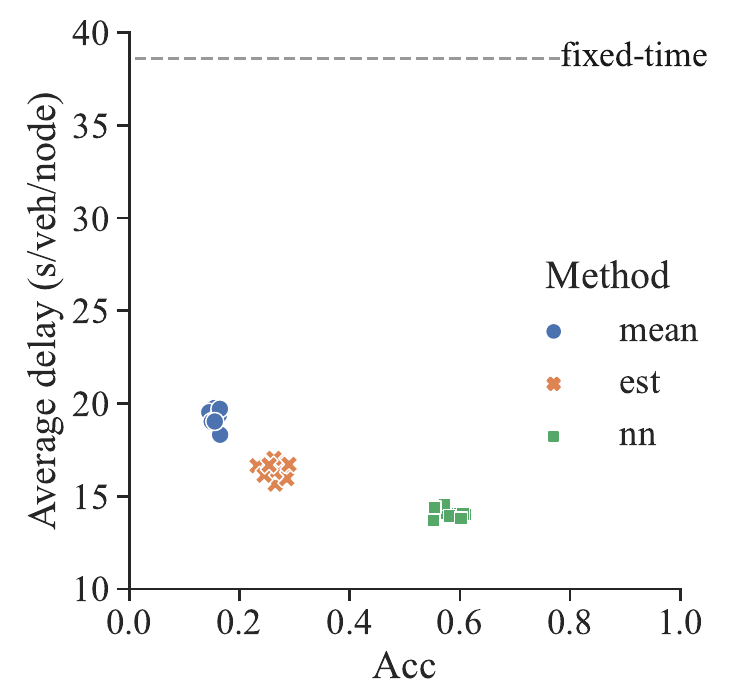}}
		\label{F:pwbp0_delay}}
	~~
	\subfloat[][LESCBP]{\resizebox{0.32\textwidth}{!}{
			\includegraphics[width=0.4\textwidth]{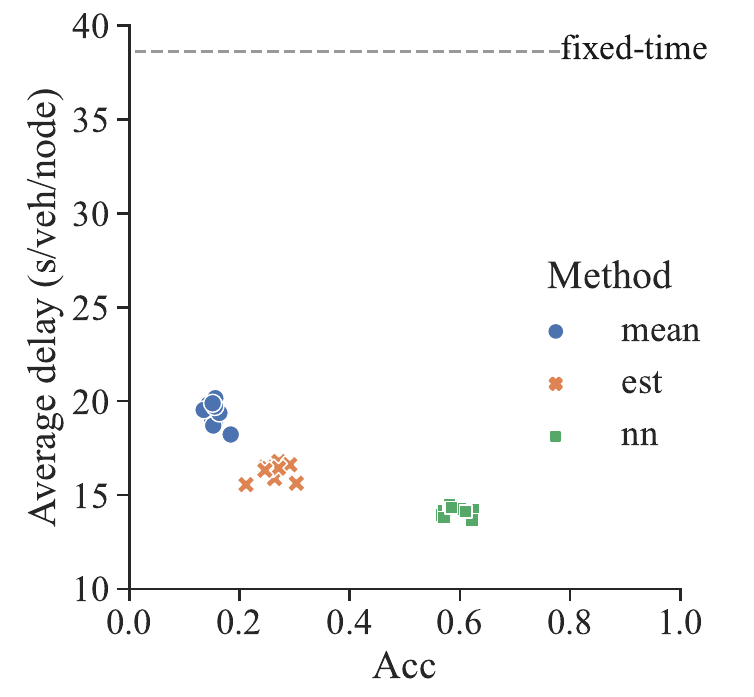}}
		\label{F:lescbp0_delay}}
	\caption{Delay for different controls with only cars.}
	\label{F:0_delay}
\end{figure}

With the same network and demand pattern in the real world ($100\%$), we replace buses and trucks with cars and exclude non-motorized vehicles. The average delays for three controls are shown in \autoref{F:0_delay}. The prediction methods remain the same (including for nn, we did not re-train it), and their prediction accuracies are still significantly different. Compared with the average delay under real-world demand, the delay with only cars decreases by about 1 to 5 seconds per node. And increasing the prediction accuracy helps to reduce vehicle delays. 

\section{Independence between prediction error and queue length}
\label{Ap:independence}
\begin{figure}[H]
	\centering
	\subfloat[][Prediction with ``mean'' method: the value and significance of Fisher exact test are 87.43 and 0.38 ($>0.05$).]
	{\resizebox{0.575\textwidth}{!}{
			\includegraphics[width=0.5\textwidth]{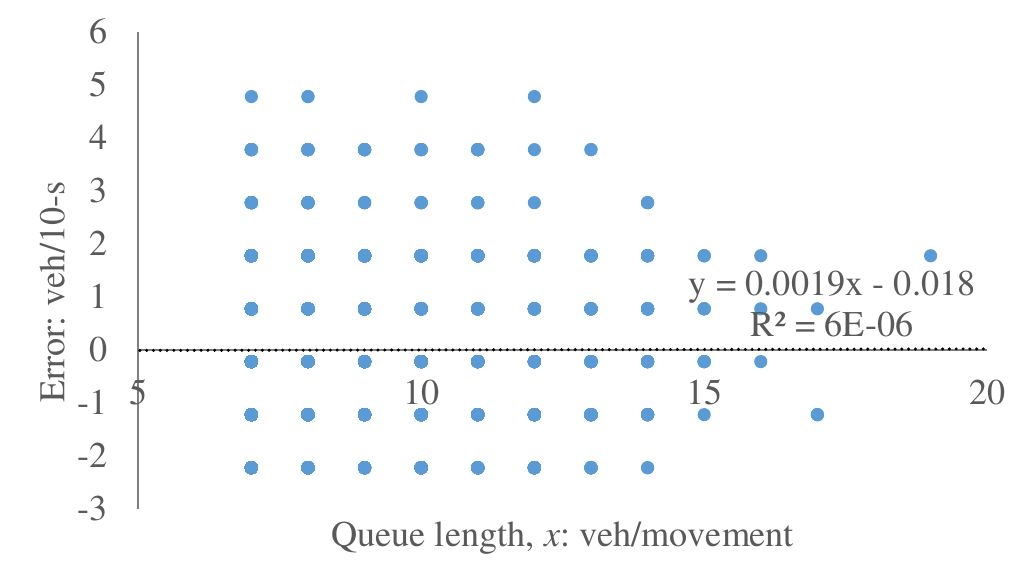}}
		\label{F:em}} 
	
	\subfloat[][Prediction with ``est'' method: the value and significance of Fisher exact test are 152.42 and 0.16 ($>0.05$).]
	{\resizebox{0.575\textwidth}{!}{
			\includegraphics[width=0.4\textwidth]{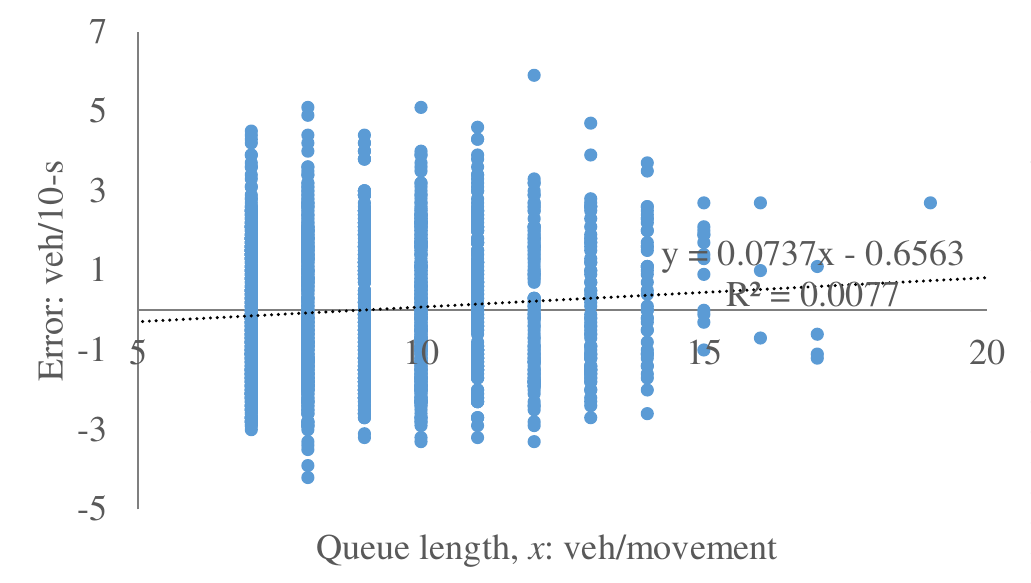}}
		\label{F:ee}}
	
	\subfloat[][Prediction with ``nn'' method: the value and significance of Fisher exact test are 105.26 and 0.55 ($>0.05$).]
	{\resizebox{0.575\textwidth}{!}{
			\includegraphics[width=0.4\textwidth]{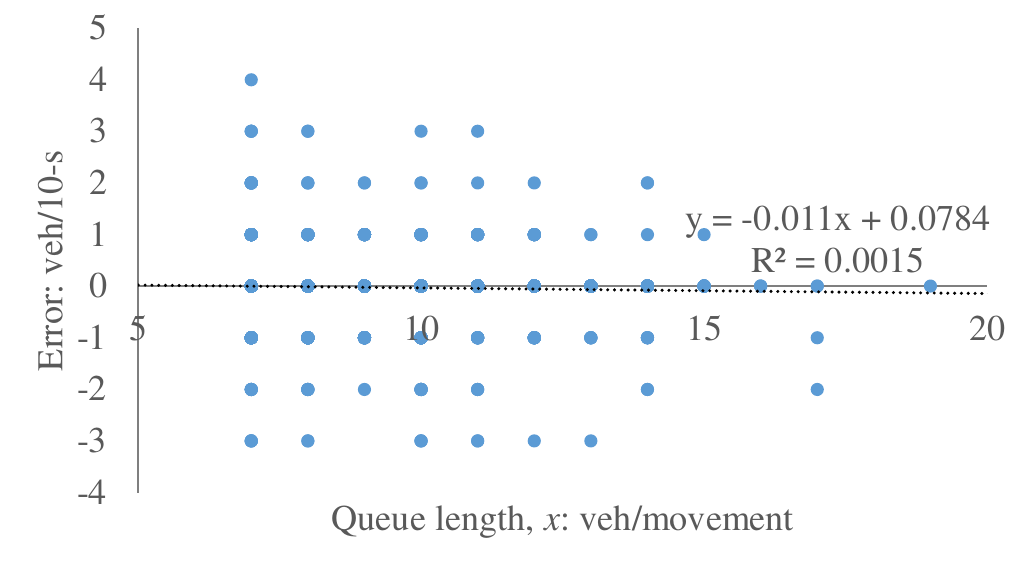}}
		\label{F:en}}
	\caption{Scatters of errors at movement { \textcircled{\smaller 1}}-north-L with 1,854 data samples.}
	\label{F:error}
\end{figure}

We collect 1854 saturated discharging samples and use three methods to predict the I-SFR, respectively. The prediction error (difference between the real I-SFR and the predicted one) and corresponding queue lengths are shown in \autoref{F:error}. Clearly, for all three methods, we cannot reject the assumption that "prediction error is independent of queue length".

In short, although the queue length information may contribute to the prediction accuracy (the occupancy information used in the ``nn'' method), it does not mean that the prediction error is not independent of queue length. This is common in multiple linear regression. Even if adding a variable can significantly reduce the deviation of prediction error, it does not mean the added variable and the prediction error are dependent.

\bibliography{ref}
\end{document}